\documentclass[aps,twocolumn,superscriptaddress,nofootinbib,prb]{revtex4-2}

\usepackage{graphicx}% Include figure files
\usepackage{amssymb,amsmath,amsfonts,bm}
\usepackage{subfigure}
\usepackage[normalem]{ulem}
\usepackage{hyperref}
\usepackage[dvipsnames]{xcolor}
\hypersetup{urlcolor=RoyalBlue, colorlinks=true,citecolor=RoyalBlue,linkcolor=RoyalBlue} 
\usepackage{physics}
\usepackage[final]{showkeys}
\usepackage{float}

\begin{document}
	\title{Flux confinement-deconfinement transition of dimer-loop models on three-dimensional bipartite lattices}
	\author{Souvik Kundu}
	\affiliation{\small{International Centre for Theoretical Sciences, Tata Institute of Fundamental Research, Bengaluru 560089, India}}
	%\affiliation{\small{Tata Institute of Fundamental Research, 1 Homi Bhabha Road, Mumbai 400005, India}}
	%\thanks{\small{Souvik Kundu was at TIFR, Mumbai when a large part of this work was done.}}
	\author{Kedar Damle}
	\affiliation{\small{Tata Institute of Fundamental Research, 1 Homi Bhabha Road, Mumbai 400005, India}}
	
	\begin{abstract}
Motivated by recent work that mapped the low-temperature properties of a class of frustrated spin $S=1$ kagome antiferromagnets with  competing exchange and single-ion anisotropies to the fully-packed limit (with each vertex touched by exactly one dimer or
  nontrivial loop) of a system of dimers and nontrivial (length $s > 2$) loops on the honeycomb lattice, we study this fully-packed dimer-loop model on
   the three-dimensional bipartite cubic and diamond lattices as a function of $w$, the relative fugacity of dimers.  We find that the 
    $w \rightarrow 0$ O($1$) loop-model limit is separated from the $w \rightarrow \infty$ dimer limit by a geometric phase transition
     at a nonzero finite critical fugacity $w_c$: The
      $w>w_c$ phase has short loops with an exponentially decaying loop-size distribution, while the $w<w_c$ phase is dominated by large loops whose loop-size
       distribution is governed by universal properties of the critical O($1$) loop soup. This transition separates two {\em distinct} Coulomb liquid phases of the system: Both phases admit a description in terms of a fluctuating
        divergence-free polarization field $P_{\mu}(\mathbf{r})$ on links of the lattice and are characterized by dipolar
         correlations at long distances. Away from the critical point, both 
          the phases are described by an effective Gaussian action in terms of the coarse-grained analog of the microscopic polarization field $P_{\mu}(\mathbf{r})$. The transition at $w_c$ is a flux confinement-deconfinement transition: For periodic samples, half-integer fluxes (along each periodic direction) of the polarization field proliferate in the
          $w < w_c$ phase, while being confined to integer values in the $w>w_c$ phase. Equivalently, and independent of boundary conditions, half-integer test charges $q=\pm 1/2$ are confined for $w>w_c$, but become deconfined in the small-$w$ phase. Although both phases are unstable to a nonzero fugacity for the charge 
          $\pm 1/2$ excitations, the destruction of the $w >w_c$ Coulomb liquid is characterized by an interesting slow crossover,
           since test charges with $q=\pm 1/2$ are confined in this phase. 
	\end{abstract}
	
	\maketitle
	\section{Introduction}
Geometrically frustrated magnets, in which the leading exchange interactions compete with each other due to the geometry of the spatial arrangement of magnetic ions in the crystal, have been and are of considerable interest to physicists. Part of the reason is that the low-energy physics in many cases is most naturally described in terms of the long-wavelength properties of a system of emergent degrees of freedom. This makes such geometrically frustrated magnets an interesting experimentally accessible arena for the study of such physics~\cite{Mila_Mendels_Lacroix_2011book,Moessner_Moore_2021book}.

A particularly interesting example of such physics is the so-called Coulomb liquid behavior of some frustrated magnets. In such systems, the low-energy and low-temperature properties are best understood in terms of an emergent divergence-free polarization field and its long-wavelength fluctuations, as well as charged excitations that violate the divergence constraint~\cite{Hermele_Balents_Fisher_2004,Banerjee_Isakov_Damle_Kim_2008,Castelnovo_Moessner_Sondhi_2008,Jaubert_Holdsworth_2009,Henley_Coulombphasereview2010,
Benton_Sikora_Shannon_2012}. In the spin-ice materials that provide perhaps the best-studied examples of such Coulomb liquids~\cite{Harris_etal_1997,Harris_Bramwell_1998,Siddharthan_Shastry_etal_1999,Bramwell_etall_2001,
Fennel_etal_2004,Fennel_etalScience2009,Castelnovo_Moessner_Sondhi_review2012,Bramwell_Harris_review2020}, a strong easy-axis single-ion anisotropy acts in consonance with the exchange and dipolar interactions between the spins to give rise to this physics when the spins form a pyrochlore lattice.

In recent work~\cite{Kundu_Damle_2025}, the present authors have identified an interesting variant of such behavior. This is predicted to arise in spin $S=1$ kagome magnets with a strong {\em easy-plane} single-ion anisotropy that {\em competes} rather than acts in consonance with comparably strong exchange interactions that act dominantly along the common out-of-plane $z$ axis. On the one-third magnetization plateau that is expected to be stabilized over a large range of magnetic fields along this axis,  the prediction is that there can be two {\em distinct} Coulomb liquid states~\cite{Kundu_Damle_2025}. Interestingly, the second Coulomb liquid can be accessed at least in some cases simply by lowering the temperature further until an unusual continuous transition is crossed~\cite{Kundu_Damle_2025}. 
Interestingly, although the critical point separating these two Coulomb liquids has some properties in common with the critical point of the two dimensional Ising model, there is no identifiable local Ising order parameter. 

Indeed, the transition is better understood as a topological phenomenon: In large-enough samples with periodic boundary conditions, the flux of the divergence-free polarization field can only take integer values along any periodic direction in one phase, while the other phase is characterized by half-integer fluxes~\cite{Kundu_Damle_2025}. An equivalent distinction, independent of boundary conditions, is that half-integer test charges are confined in one phase, but become deconfined in the other phase~\cite{Kundu_Damle_2025}. Motivated by the more well-studied physics of bipartite dimer models and the fact that the corresponding fluxes can only take on integer values in such dimer models, Ref.~\cite{Kundu_Damle_2025} dubbed this a ``flux-fractionalization'' phenomenon.
However, this terminology gives primacy to the physics of one of the two phases, and it is perhaps more appropriate to treat both sides of the transition on an equal footing and refer to this as a flux confinement-deconfinement transition. This is the convention we adopt henceforth.

For such kagome magnets, an intuitively accessible geometric characterization of this physics is in terms of the statistical mechanics of a system of dimers and nontrivial (length $s>2$) loops on the honeycomb lattice whose vertices represent the centers of the kagome triangles and whose edges host the spins of the original kagome magnet~\cite{Kundu_Damle_2025}. This dimer-loop model features a full-packing constraint requiring that each vertex of the honeycomb lattice be touched by exactly one dimer or nontrivial loop, and the relative fugacity $w$ of dimers is given by $\exp(-\mu/T)$, where $\mu \equiv \Delta - J$ is a material parameter that encodes the relatively small ({\em i.e.}, $\mu \ll \Delta,J$) difference in the strengths of the dominant easy-axis exchange coupling $J$ and the comparably strong easy-plane single-ion anisotropy $\Delta$ that competes with it. Making the natural identification between dimers and ``trivial'' length $s=2$ loops, the transition between the two Coulomb phases corresponds in this language to a transition between a short-loop phase and a critical quasi-long-loop phase of this fully-packed O(1) loop model~\cite{Kundu_Damle_2025}. The physics of this dimer-loop model was also shown to generalize naturally to the corresponding model on the square lattice~\cite{Kundu_Damle_2025}.
	\begin{figure} [t]
		\centering
		\includegraphics[width=\linewidth]{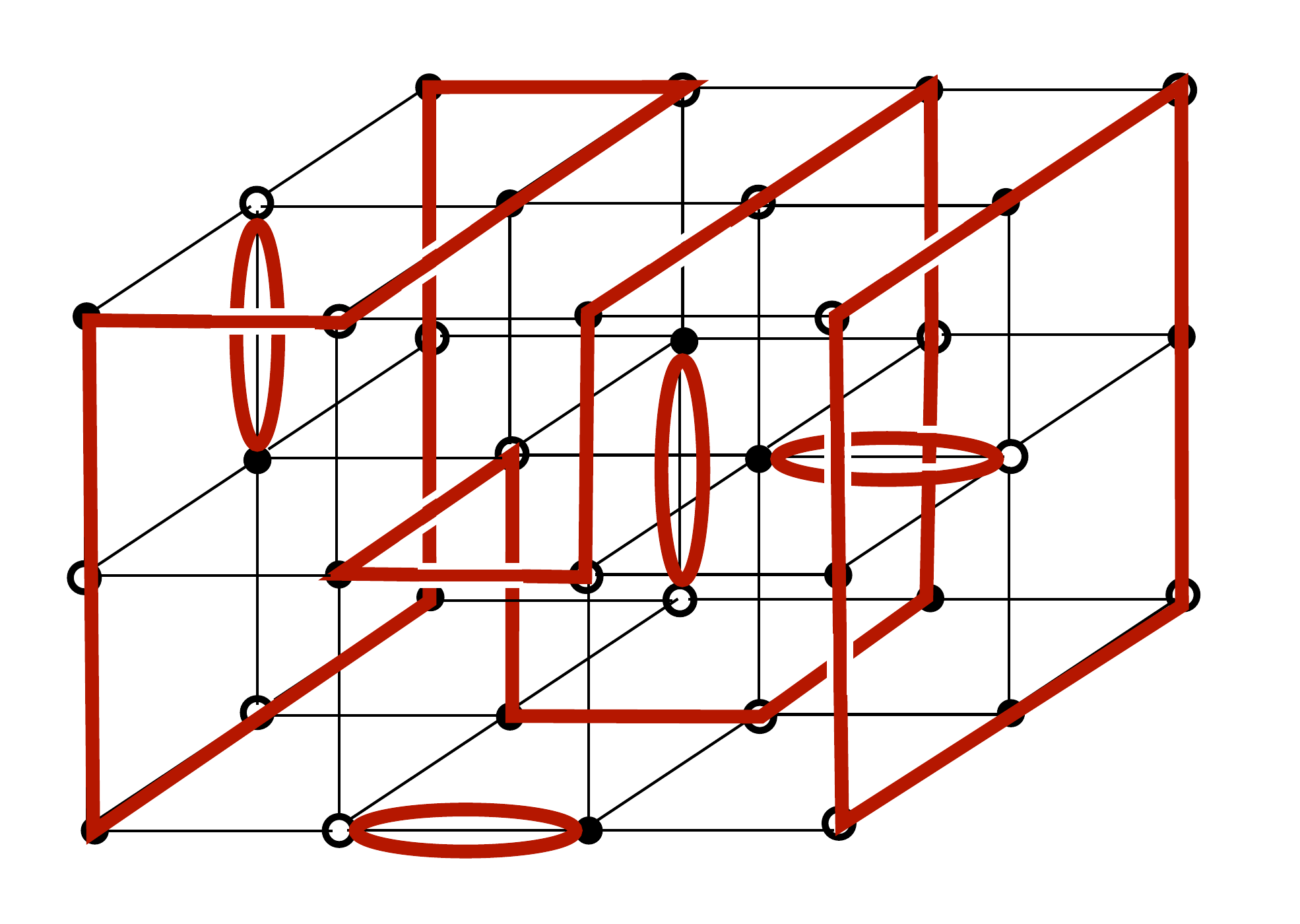}
		\caption{\label{fig:fplcubic} An example of a valid fully-packed configuration of the dimer-loop model on a $4\times3\times3$ cubic lattice with open boundary conditions. Note that the representation of dimers emphasizes their equivalence to {\em trivial} length $s=2$ loops  that traverse a single link of the lattice in both directions. In such a fully-packed configuration, each site is touched by a single loop, which can be either a dimer (trivial loop) or a nontrivial loop of length $s>2$.}
		
	\end{figure}

Since both the $w=0$ limit of fully-packed O(1) loops and the $w \rightarrow \infty$ limit of fully-packed dimers have a long history of study in statistical mechanics~\cite{Kastelyn_1961,Fisher_1961,Fisher_Stephenson_1963,Samuel_1980_1,Samuel_1980_2,Youngblood_Axe_McCoy_1980,
Youngblood_Axe_1981,Baxter_book_1989,Blote_Nienhuis_1994,Batchelor_Blotte_Nienhuis_Yung_1996,Huse_Krauth_Moessner_Sondhi_2003,
Liu_Deng_Garoni_Blote_2012,Fu_Guo_Blotte_2013,Nahum_Chalker_Serna_etal_2013_PRB,Nahum_Chalker_Serna_etal_2013_PRL}, this fully-packed dimer-loop model also represents an interesting and natural generalization of these paradigmatic models that have attracted so much interest through the years. Clearly, this rationale is not special to two dimensions. Indeed, from the point of view of statistical mechanics, the generalization to three dimensions is equally, if not more, interesting; since the bipartite O($1$) loop model in three dimensions has a genuine long-loop phase as opposed to the critical quasi-long-loop phase it exhibits in two dimensions.

	\begin{figure*} [t]
		\centering
		\includegraphics[width=\textwidth]{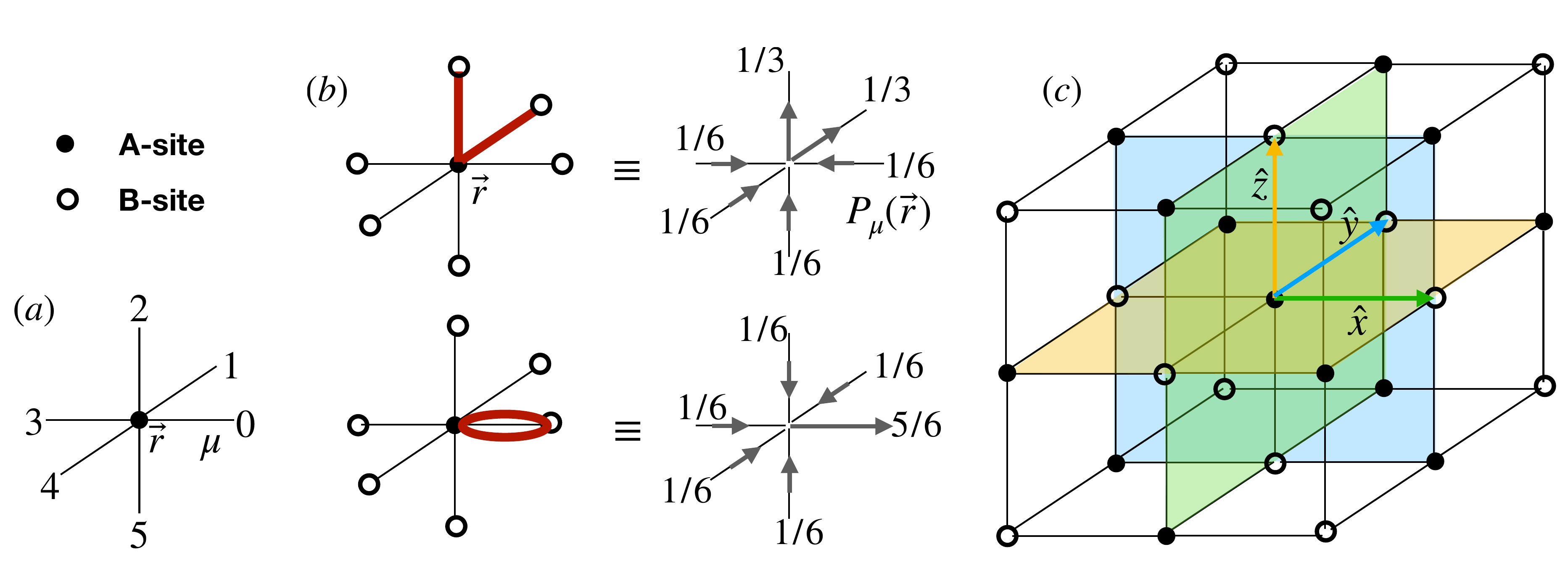}
		\caption{\label{fig:Pmucubic} ($a$) Links connecting a site to its neighbors are represented as ($\vec{r},\mu$) where $\mu$ is the orientations of the link and $\vec{r}$ the coordinate of the site.  ($b$) Two examples of the local configuration at a site---a nontrivial loop passing through the site and a trivial loop touching the site---are mapped to the corresponding polarization field $P_{\mu}(\vec{r})$ defined on the lattice links (Eq.~\ref{eq:Pmu}). In both cases, net divergence of this polarization field at $\vec{r}$ is $0$. ($c$). The three orientations of the principal planes through which we define the flux of the polarization field on the cubic lattice are shown in three colors and labeled by the direction of the normal to these planes. The construction of the polarization field and the definition of the fluxes for the diamond lattice are entirely analogous, except that the normals to the three principal planes are not orthogonal to each other, but are chosen to be along the three principal directions of the diamond lattice, along which periodic boundary conditions are imposed in our computational work. See Sec.~\ref{subsec:Pmu}, Sec.~\ref{subsec:fluxandloopsize}, and  Sec.~\ref{subsec:fluxdistribution} for details.}
		
	\end{figure*}
	
The O($1$) loop model  and the dimer model on regular bipartite lattices such as the cubic and the diamond lattice both admit a lattice-level description in terms of a fluctuating divergence-free polarization field (with violations of the full-packing constraint playing the role of charges that are sources of this polarization field), do these two different models control two distinct Coulomb phases of the dimer-loop model, or do they represent the two endpoints of a smooth crossover along the $w$ axis? If there are two distinct phases, is there any local observable whose correlations can distinguish between them?

With this motivation, we study this dimer-loop model on the cubic and diamond lattices using insights from the effective field theory description of Coulomb physics and large-scale Monte Carlo simulations. We find that there are two distinct Coulomb phases separated by a phase transition at a nonzero finite critical coupling $w_c$ on both lattices. The $w>w_c$ phase has short loops and dimers, whereas the $w<w_c$ phase is dominated by extended loops, with the largest loop in a sample of linear size $L$ scaling as $s_{\rm max} \sim L^3$ and the distribution of large loops obeying a universal Poisson-Dirichlet (PD) distribution that has been argued earlier to describe the properties of the O($1$) loop model in three dimensions~\cite{Nahum_Chalker_Serna_etal_2013_PRL}.  Both phases admit a unified description in terms of a Gaussian effective action for the coarse-grained version of the lattice-level polarization field. 

In terms of their Coulomb physics, they are distinguished by the fact that half-integer fluxes proliferate in the $w< w_c$ phase, but are confined to integer values in the large-$w$ phase in samples with periodic boundary conditions. Equivalently, and independent of boundary conditions, the two phases are distinguished by the behavior of half-integer test charges of $q = \pm 1/2$.  In the flux-fractionalized phase ($w<w_c$), two test charges with half-integer values are deconfined. This means they can be separated by any arbitrary distance with only a finite energy cost. In contrast, in the $w>w_c$ phase, these half-integer charges are tightly bound into integer-valued objects which are in turn deconfined.

This distinction has consequences for the nature of the instability associated with a nonzero fugacity $f_{1/2}$ for the elementary $q \ \pm 1/2$ charges. Both the $w<w_c$ flux-fractionalized Coulomb phase and the $w>w_c$ Coulomb phase are unstable in the presence of a nonzero $f_{1/2}$. However, in the $w>w_c$ phase, the instability takes the form of a very slow crossover. Heuristically, we understand this as follows: Since half-integer charges are confined in the $w>w_c$ phase, they cannot proliferate without first binding into integer-charged objects. Very schematically, in renormalization-group terms, $f_{1/2}$ is expected to be marginally relevant in the $w>w_c$ phase, in the sense that a nonzero $f_{1/2}$ needs to first induce a nonzero value of the $f_{1}$, the fugacity for integer charges, and it is this nonzero $f_{1}$ that is a relevant perturbation.
This is different from the flux-fractionalized phase in which a nonzero $f_{1/2}$ is a relevant perturbation (as opposed to a marginally relevant perturbation).

	The rest of the paper is organized as follows: In Sec.~\ref{sec:model} we describe the fully-packed dimer-loop model and its descendants that include defects, as well as the lattice-level mapping to a divergence-free polarization field. In Sec.~\ref{sec:methodsandobservables3d}, we define all the observables of interest,  and describe a generalization (to include defects) of the half-vortex worm algorithm described in Ref.~\cite{Kundu_Damle_2025} in the context of fully-packed dimer-loop model.  We present our numerical results in Sec.~\ref{sec:results} and we conclude with a discussion in Sec.~\ref{sec:Discussion}.

	\section{The dimer-loop model and its extensions}
	\label{sec:model}

	The fully-packed dimer-loop model introduced in Ref.~\cite{Kundu_Damle_2025} is defined as follows: Each vertex of a lattice is constrained to be touched or visited by exactly one simple loop whose segments live on links of the lattice. Such a loop can either be a ``nontrivial'' loop with length $s > 2$, or a ``trivial'' loop of length $s=2$ that traverses a single link of the lattice in both directions; naturally, the latter can be identified with a hard-core dimer on this link. Each trivial loop of length $s=2$ has fugacity $w$, while nontrivial loops of even length $s>2$ are assigned unit fugacity. An example of a fully-packed dimer-loop configuration on the cubic lattice is shown in Fig.~\ref{fig:fplcubic}.  
	
	The weight $W(\mathcal{C})$ of a valid configuration $\mathcal{C}$ ({\em i.e.}, obeying the full-packing constraint) is given by $W(\mathcal{C}) = w^{n_d(\mathcal{C})}$, where $n_d(\mathcal{C})$ is the number of trivial loops or dimers in a valid configuration. 	On any lattice in which local rearrangements at the level of a single plaquette are possible without violating the full-packing and hard-core constraints, there are an exponentially large (in system volume) number of configurations ${\mathcal C}$, and there is a nonzero entropy density at any $w$. The corresponding partition function of this model is:
	\begin{equation}\label{eq:Zfpl}
		Z(w) = \sum_{\mathcal{C}}w^{n_d(\mathcal{C})}.
	\end{equation}

	\begin{figure} [t]
		\centering
		\includegraphics[width=\linewidth]{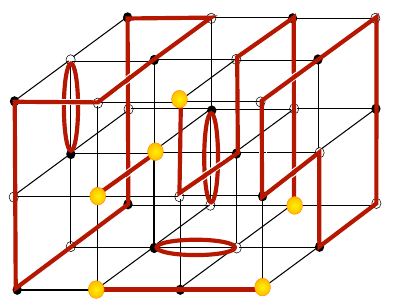}
		\caption{\label{fig:nonfplcubic} When the fugacity $f_{1/2}$ for half-charge defects is nonzero, the partition function of the generalized dimer-loop model has contributions from configurations that are not fully packed. An example of such a configuration is shown here for a $4\times3\times3$ cubic lattice with open boundary conditions. Sites colored yellow are touched by a single segment of a nontrivial loop, and thus host half-charge defects, {\em i.e.}, the corresponding configuration of the polarization field has a nonzero divergence at these vertices, corresponding to a charge $q=\pm 1/2$ located at these vertices. These defects always come in pair which terminates an open string of arbitrary length, including length $s=2$ open strings. There are thus two kinds of objects that occupy a single link of the lattice: trivial loops (dimers) of loop length $s=2$ that traverse a single link in both directions, and open strings of length $s=2$. }
	\end{figure}

	\subsection{Mapping to divergence-free polarization field}\label{subsec:Pmu}
	We define $n_{\mu}(\mathbf{r})$ to be the occupation number of a link
	 $(\mathbf{r},\mu)$ 
	 (here we choose a standard convention whereby $\mu$ is the direction of the link, and $\mathbf{r}$ the vertex at one of its ends). 
	 A link occupied by a dimer has $n_{\mu}(\mathbf{r}) =1$, while a link occupied by a segment of a nontrivial loop has $n_{\mu}(\mathbf{r}) =1/2$, and an empty link has $n_{\mu}(\mathbf{r}) =0$.
	On a regular bipartite lattice, we then have the following lattice-level mapping to a polarization field vector $P_{\mu}(\mathbf{r})$,
	\begin{equation}\label{eq:Pmu}
		P_{\mu}(\mathbf{r})|_{A\to B} = n_{\mu}(\mathbf{r})-1/z
	\end{equation}
	Here, $z$ is the lattice coordination number and the subscript $|_{A\to B}$ denotes the fact that the right-hand side gives the component of the polarization field directed along the link $(\mathbf{r},\mu)$, from its $A$-sublattice endpoint to its $B$-sublattice endpoint.  
	
	The full-packing and hard-core constraints now imply that the lattice divergence of $P_{\mu}(\mathbf{r})$  is zero, {\em i.e.}, $\mathbf{\Delta} \cdot \mathbf{P} = 0$ (see Fig.~\ref{fig:Pmucubic}) except on the boundary of a sample with open boundary conditions. Here, we work with periodic boundary conditions, and this polarization field is divergence-free throughout the sample.
	Since this is the case, the corresponding flux $\phi_{\mu}$ in the direction $\mu$, perpendicular to any principal plane $\sigma_{\mu}$ is well-defined and independent of choice of plane used in its computation. Thus, we may choose an arbitrary plane $\sigma_{\mu}$ with normal along $\mu$ and compute this flux as a sum of polarization fluxes on the set of links $(\mathbf{r},\mu)_{\sigma}$ piercing this plane:
	\begin{eqnarray}
		\phi_{\mu} = \int_{\sigma}\mathbf{P}\cdot{\rm d}\bm{\sigma} = \sum_{(\mathbf{r},\mu)_{\sigma}}P_{\mu}(\mathbf{r}).
	\end{eqnarray}

With periodic boundary conditions, this flux $\phi_{\mu}$ along each periodic direction $\mu$ is a topological property of the configuration in the sense that it cannot be altered by local changes to the configuration. Further, it is easy to see that it is constrained at the microscopic level to take on half-integer values, except in the $w = \infty$ limit of fully-packed dimers, where $\phi_{\mu}$ can only take on integer values. 
Finally, note that one can solve the divergence-free constraint on a three-dimensional bipartite lattice by writing this divergence-free polarization field $P_{\mu}(\mathbf{r})$ as the lattice curl $\mathbf{P} = \nabla\times\mathbf{A}$  of a lattice-level vector potential $A_{\mu}$ defined on the links of the dual lattice. A coarse-grained version of this representation plays a central role in the theory of the Coulomb correlations of fully-packed dimer models on bipartite three-dimensional lattices, which models these correlations via an effective action of the form $S_{\rm eff} = \frac{K}{2}\int  d^3x \mathbf{P}^2$~\cite{Huse_Krauth_Moessner_Sondhi_2003}. By an extension of these ideas, one also expects that the fully-packed loop model on three-dimensional bipartite lattices has a similar coarse-grained description. As will be clear from the discussion in Sec.~\ref{subsec:fluxdistribution}, such an effective field theory also proves useful in understanding the long-distance properties of the general dimer-loop model as a function of $w$.

	\subsection{Generalization to include half-vortices}\label{subsec:extendedmodel}
If one removes a single segment of a nontrivial loop, one creates an open string whose free ends host charges $\pm 1/2$ that serve as sources of the polarization field and violate its divergence-free condition at these two ends. Once such an open string has been created, subsequent local changes in the configuration can of course move these charges apart. Likewise, if a vertex is not touched by a dimer or visited by a nontrivial loop, it hosts a unit $\pm 1$ charge. When the fugacities $f_{1/2}$ and $f_{1}$ for such charges are nonzero, the fluxes $\phi_{\mu}$ are of course no longer well-defined. In the dimer-loop model, the fundamental defects are pairs of $\pm 1/2$ charges, which we will also refer to loosely as vortices, to connect with the terminology used in Ref.~\cite{Kundu_Damle_2025}. This is because integer $\pm 1$ charges can be thought of as bound states of these basic half-integer defects.
	
	One can extend the fully-packed dimer-loop model to include a finite density of half-integer vortices by introducing a nonzero fugacity $f_{1/2}$ for the half-integer charged vortices. This leads to a generalized dimer-loop-string model, where both open strings of arbitrary length ($s \geq 2$) and closed loops coexist. However, the fugacity $f_{1}$ of unit charges continues to be set to zero, so that every vertex is either visited by a nontrivial loop, or touched by a dimer, or has an open string that terminates on it. Notice that in this generalized model, there are two quite distinct objects of length $s=2$: open strings of length $s=2$, and trivial loops (dimers). The distinction is that the latter carry charges $\pm 1/2$ at their two ends, while a dimer is a trivial loop of length $s=2$.
Thus, this modification introduces a rich variety of additional configurations, and, potentially, new physics. An example of such a configuration is shown in Fig.~\ref{fig:nonfplcubic}. 
	
	The Boltzmann weight $W(\mathcal{C}^{'})$ of a configuration $\mathcal{C}^{'}$ containing $n_d(\mathcal{C}^{'})$ dimers and $n_o(\mathcal{C}^{'})$ open strings is now given by $W(\mathcal{C}^{'}) = w^{n_d(\mathcal{C}^{'})} f_{1/2}^{2n_o(\mathcal{C}^{'})}$. Here $w$ is the weight per dimer and $f_{1/2}$ controls the density of half-integer vortices. The corresponding partition function of this extended model is:
	\begin{equation}\label{eq:Znonfpl}
		Z(w,f_{1/2}) = \sum_{\mathcal{C}^{'}} w^{n_d(\mathcal{C}^{'})}.f_{1/2}^{2n_o(\mathcal{C}^{'})} = \sum_{\mathcal{C}^{'}} w^{n_d(\mathcal{C}^{'})}.f_{1/2}^{n_v(\mathcal{C}^{'})},
	\end{equation}
	where $n_v(\mathcal{C}^{'})$ denotes the number of half-integer vortices.
	
	In the context of spin $S=1$ kagome lattice antiferromagnet explored in Ref.~\cite{Kundu_Damle_2025}, the magnetic field-driven transition from the $1/3$ magnetization plateau to the $2/3$ magnetization plateau of the kagome lattice can be mapped to this generalized dimer-loop-string model on the honeycomb lattice; this was the original motivation of Ref.~\cite{Kundu_Damle_2025} for generalizing the fully-packed dimer-loop model to include half-integer vortices.  We emphasize, however, that it is not entirely clear if such a dimer-loop-string model on the diamond lattice can describe the physics of any frustrated spin $S=1$ magnet on the pyrochlore lattice. This is because the local anisotropy axes at the pyrochlore sites are not expected to be collinear to each other in such spin systems; this is briefly discussed further in Sec.~\ref{sec:Discussion}.

	\subsection{Focus of present study}\label{subsec:definition}	
	
 Motivated by the connection to the low-temperature physics of a class of spin $S=1$ kagome magnets, Ref.~\cite{Kundu_Damle_2025} focused on a detailed study of the phase diagram of the fully-packed dimer-loop model as a function of $w$ in two dimensions, specifically on the honeycomb and square lattices, and identified the unusual flux confinement-deconfinement transition alluded to in the Introduction. 	
	What aspects of this two-dimensional physics generalize to the three-dimensional case? Motivated by this question, we study here the phase diagram of the fully-packed system, with partition function $Z(w)$, on the cubic and diamond lattices. 
	
	Ref.~\cite{Kundu_Damle_2025} also argued that the instability induced by a nonzero $f_{1/2}$ is very different in the two Coulomb phases that are separated by this flux confinement-deconfinement transition. More precisely, Ref.~\cite{Kundu_Damle_2025} argued that charge $\pm 1/2$ defects were 
	in themselves an irrelevant perturbation in the large-$w$ Coulomb phase, while charge $\pm 1$ defects, corresponding to
	 vertices not visited by any loop at all, were strongly relevant perturbations in this phase. Since half-integer
	  charges can form bound states that have integer charge, Ref.~\cite{Kundu_Damle_2025} further argued that a nonzero fugacity $f_{1/2}$ for half-integer defects would destabilize the large-$w$ Coulomb phase, even when $f_{1}$--the fugacity for integer-charged defects--was set to zero. However, this instability would present itself as a very slow crossover. This behavior was expected to be very different from the corresponding instability of the small-$w$ flux-fractionalized phase, since defects with half-integer charge are themselves a strongly relevant perturbation in the flux-fractionalized phase. 
  
	  Our second goal in this study is to test these ideas by extending the two-dimensional computations of Ref.~\cite{Kundu_Damle_2025} to the case of nonzero $f_{1/2}$. Specifically, we study the dimer-loop model with a small nonzero $f_{1/2}$ on the square lattice, and characterize the difference in the associated instabilities of both the Coulomb phases of this system.

	\section{Methods and observables}	
	\label{sec:methodsandobservables3d}
	We have employed classical Monte Carlo simulations, utilizing two variants of the worm algorithm~\cite{Sandvik_Moessner_2006, Alet_Ikhlef_Jacobsen_etal_2006}---the half-vortex worm algorithm and the unit-vortex worm algorithm---to perform large-scale simulations on cubic and diamond lattices. For the case of the fully-packed dimer-loop model on any bipartite lattice, Ref.~\cite{Kundu_Damle_2025} already provides a detailed account of these variants, which are both based on the ``myopic'' methodology developed in Ref.~\cite{Rakala_Damle_2017} for guaranteeing detailed balance in a particularly simple and transparent way. 
	
	Here we generalize these to include a nonzero $f_{1/2}$. Our generalization borrows from a similar use of the myopic methodology~\cite{Rakala_Damle_2017} in Ref.~\cite{Morita_Lee_Damle_Kawashima_2023} to obtain an efficient grand-canonical algorithm for the dimer-loop model with nonzero fugacity of half-integer vortices.  This generalization is also broadly applicable for equilibrium sampling of constrained statistical systems like spin-ice states and their low-energy charge excitations of higher spin systems, and expected to be useful in a variety of contexts; this is also discussed briefly below.	
	\begin{figure}[t]
		\centering
		\includegraphics[width=\linewidth]{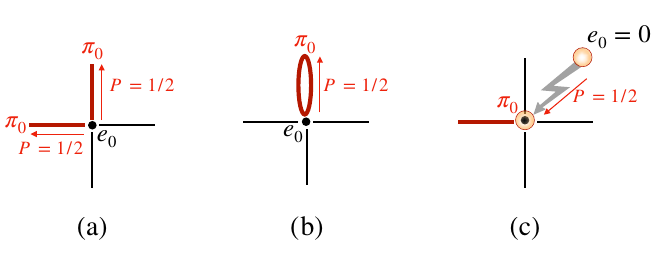}
		\caption{\label{fig:initentry} a), b), c) Inequivalent local environments of a square lattice site. In general, a worm starts with a random site (the black circle) which can have one of these local environments. In the first two cases (a,b) the worm tail is fixed at this site, which becomes the entry site $e_0$ for the first pivot encountered in the worm construction. In (a) this first pivot is chosen randomly from the two possibilities displayed, with a probability $1/2$ for each. In (b), the choice of this pivot $\pi_0$ is uniquely determined by the orientation of the dimer, but is used only with a probability $1/2$, while the worm construction is aborted without doing anything with probability $1/2$. In (c),  if the randomly placed site is associated with a half-integer charge, the worm update is either aborted with probability $1/2$ without doing anything, or this site is itself chosen as the first pivot with probability $1/2$. In the latter case, the entry to this first pivot is said to be from an ``off-lattice'' entry point $e_0=0$.}
		
	\end{figure}
	
				\begin{figure}[t]
		\centering
		\includegraphics[width=\linewidth]{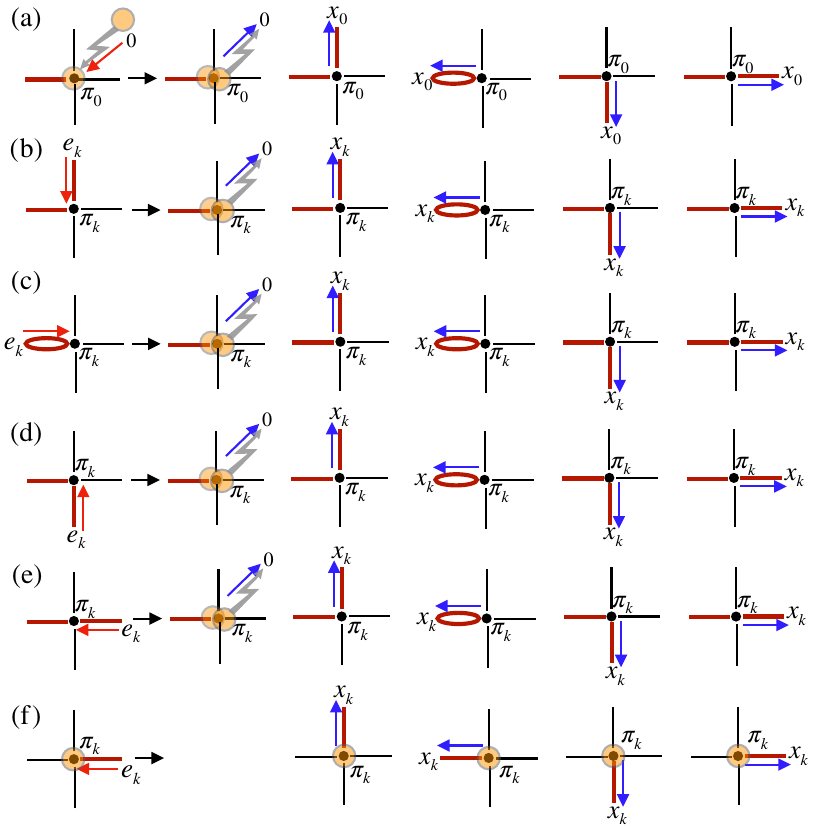}
		\caption{\label{fig:exitleg} Given that a pivot was entered from entry $e_k$, the exit $x_k$ via which the pivot is exited by the worm head is chosen from the allowed possibilities using probabilities assigned to each choice. (a) ---(f) This figure provides a pictorial illustration of the local information required to determine the relative weights (Eq.~\ref{eq:weight}) that appears in detailed balance equation set (Eq.~\ref{eq:detailed_balance}) that these probabilities must satisfy. Thus, the ``off-lattice" entrance/exit choices are depicted with two half-integer charges at the position of the pivot. This is consistent with the fact that a factor of $f_{1/2}^2$ is incorporated in the corresponding relative weights. In case (d) the off-lattice exit is forbidden since that would lead to a unit-vortex of vorticity $\pm 1$ at $\pi_k$, whose fugacity $f_{1}$ has been set to zero in the present study. }
	\end{figure}

	\begin{figure}[t]
		\centering
		\includegraphics[width=\linewidth]{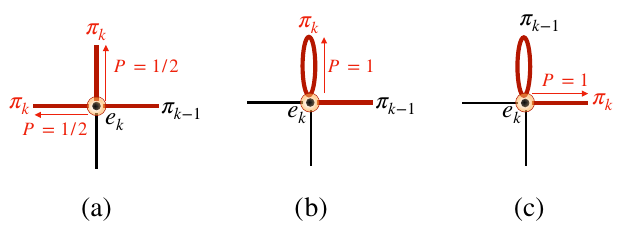}
		\caption{\label{fig:choosepivot} After an exit $x_{k-1}$ has been chosen at the previous pivot $\pi_{k-1}$ encountered during the worm construction, the worm head moves to this exit site if this exit is not the ``off-lattice exit''. This exit site becomes the entry site $e_k$ for the next pivot $\pi_k$. This next pivot is chosen ``myopically'', {\em i.e.}, without regard to the relative Boltzmann weights associated with the various choices. The manner in which this is done depends on the local configuration around $x_{k-1} \equiv e_k$. Apart from the loop segment or dimer that connects $x_{k-1}$ to $\pi_{k-1}$, $x_{k-1} \equiv e_k$ can be touched by (a) either two loop segments or (b) one dimer (if $x_{k-1}$ is connected to $\pi_{k-1}$ by a loop segment) or (c) it is necessarily touched by a single loop segment (if the link connecting $x_{k-1}$ and $\pi_{k-1}$ has a dimer on it after the pivot $\pi_{k-1}$ is exited). In the former case (a), $\pi_k$ is chosen with probability $1/2$ each to be one of these two sites connected to $e_k$ by these two loop segments, while in (b), it is chosen deterministically to be the site connected to $e_k$ by this dimer. In the latter case (c), the pivot $\pi_k$ is deterministically chosen to the site connected to $e_k$ by this single loop segment. }
	\end{figure}
	
		\subsection{Generalized half-vortex worm algorithm}\label{subsec:generalizedMC}
	The half-vortex worm starts from a random site $v_0$ of the lattice. Two possibilities exist (See Fig.~\ref{fig:initentry}): First, $v_0$ may initially host a defect of charge $\pm 1/2$. Second, $v_0$ initially has no charge associated with it. We describe the procedure to be followed for each case in turn.

	In the first case, we either abort the worm construction with probability $1/2$ or we choose $v_0$ as the first ``pivot'' site $\pi_0$ with probability $1/2$. If the latter choice is made, we view the pivot $\pi_0$ as being reached from an ``entry'' site $e_0=0$ that is ``outside of the lattice" (see Fig.~\ref{fig:initentry} (c)). In the second case, in which $v_0$ initially had no charge associated with it, this initial site will be the first ``entry site'' $e_0$ where the ``worm tail'' will be held fixed for the duration of the worm construction. In this case, if $v_0$ is connected to a neighboring site $v_1$ via a dimer (Fig.~\ref{fig:initentry} (b)), we either abort the worm with probability $1/2$ or we choose $v_1$ to be the first pivot site $\pi_0$ with probability $1/2$. If $v_0$ is connected to two neighboring sites via two segments of a nontrivial loop (Fig.~\ref{fig:initentry} (a)), we randomly select one of these neighbors to be the first pivot site $\pi_0$ with equal probability $1/2$ each. In either event, we have now ``entered'' this first pivot site via the entry site $e_0$. We view the pivot site as the position of the ``worm head'', and the algorithm can be thought of as the construction of a directed walk that is undertaken by the worm head while keeping the tail fixed.
	
In general, there are $z+1$ possibilities for ``exiting" from a pivot, in this case $\pi_0$. We denote the chosen exit as $x_0$. The respective probabilities for using each of these possible exits are chosen to satisfy local detailed balance. To efficiently obtain these probabilities, a table is constructed and stored in advance; we will discuss the construction of this probability table below. Independent of the choice of exit, we first delete the loop segment hosted by the link connecting $\pi_0$ to $e_0$ as the prerequisite for executing an eventual exit. If the link connecting $\pi_0$ to $e_0$ originally hosts a dimer (a nontrivial loop of length $s=2$), ``deleting the loop segment'' is understood to mean that this dimer is in fact converted into a loop segment.

In our numbering convention, the first $z$ of these possible exits correspond to choosing a neighbor as the exit site $x_0$, and placing one loop segment on the link connecting $\pi_0$ to $x_0$.  If the link in question already has a loop segment on it, placing another loop segment on it as described above is understood to convert the original loop segment into a dimer, {\em i.e.}, a trivial loop of length $s=2$.  If one of these $z$ exits is chosen, this neighbor will either serve as the ``entry'' to the next pivot, or serve as the termination point of the worm construction; the manner in which these choices are made is described later.  

The $z+1$'th exit option corresponds to exiting via an ``off-lattice exit". For the special case of the first pivot $\pi_0$ that hosts a defect of charge $\pm 1/2$, choosing the off-lattice exit simply means  aborting the worm without doing anything. For the case of the initial pivot reached from a defect-free entry site, choosing the off-lattice exit creates a pair of $\pm 1/2$ charge defects at $e_0$ and $\pi_0$, since we always remove a loop segment from the link connecting $\pi_0$ to $e_0$ as a prelude to choosing an exit. Note that this off-lattice exit is only allowed if $\pi_0$ did not originally host a half-charge defect before the worm head reached this pivot. Otherwise, choosing the $z+1$'th off-lattice exit would convert this half-charge defect into a unit-charge defect, which has zero fugacity and is therefore forbidden.

 Likewise, after reaching any subsequent pivot $\pi_k$ via an entry site $e_k$, choosing the $z+1$'th exit option corresponds more generally as well to ending the worm construction after removing the loop segment connecting $e_k$ to $\pi_k$ to create a charge $\pm 1/2$ defect at $\pi_k$. As before, this is only allowed if $\pi_k$ did not originally host a half-charge defect before the worm head reached this pivot. 
 %Otherwise, choosing the $z+1$'th off-lattice exit would convert this half-charge defect into a unit-charge defect, which has zero fugacity and is therefore forbidden.
 
 Note that at such subsequent pivots, this does not create a half-charge defect at $e_k$. Indeed, if the worm move had started at a site $v_0$ that was originally free of any defect, the other half-charge defect is being held fixed with the worm tail at $v_0$ throughout the worm construction.
Similarly, at any subsequent pivot $\pi_k$ reached from entry site $e_k$,  choosing one of the other $z$ exits, say $x_k$, corresponds to removing the loop segment that originally connected $\pi_k$ to $e_k$, and placing it instead on the link connecting $\pi_k$ to $x_k$. Again, if the link connecting $\pi_k$ to $e_k$ originally hosted a dimer, this converts it into a loop segment. And if the link connecting $\pi_k$ to $x_k$ originally had a loop segment on it, this converts this loop segment into a dimer. Naturally, if $x_k$ coincides with $e_k$, we bounce back to $e_k$ without any change in the local configuration. All of this is summarized pictorially in Fig.~\ref{fig:exitleg}.

 Next, we describe the procedure to be followed at an exit site $x_k$:
If this exit site originally hosted a charge $\mp 1/2$ defect, this defect is now healed by the placement of a loop segment on the link connecting $x_k$ to $\pi_k$ and the worm construction ends. In this case, one has either removed a pair of half-charge defects at $e_0$ and $x_k$ (if the worm construction had started at a site $v_0$ that originally had a charge $\pm 1/2$ on it), or one has moved the charge $\mp 1/2$ defect from $x_k$ to $v_0$ (if the worm construction had started at a site $v_0$ that originaly had no defect on it).

On the other hand, if the exit $x_k$ was originally defect-free, then it was already touched either by a dimer that connects it to a neighbor $v_d$, or by a pair of loop segments connecting it to two neighbors $v_n$ and $v_n'$. In the former case, the worm head moves to $v_d$, which becomes the next pivot $\pi_{k+1}$. The latter case splits into two subcases: either one of these neighbors $v_n$ and $v_n'$ coincides with $\pi_k$, or they are different from $\pi_k$. If they are different from $\pi_k$, then the worm head moves to either $v_n$ or $v_n'$ with probability $1/2$ each, and the chosen neighbor becomes the next pivot $\pi_{k+1}$. Whereas if $\pi_k$ coincides with, say, $v_n$, then the worm head moves to $v_n'$ with probability $1$, and it becomes the location of the next pivot $\pi_{k+1}$.  In all cases, $x_k$ is now identified as the entry site $e_{k+1}$ from the viewpoint of the next pivot $\pi_{k+1}$. All of this is summarized in Fig.~\ref{fig:choosepivot}.

It only remains to describe the construction of the probability table governing the choice of exit at a pivot $\pi_k$. For this purpose, we consider the local detailed balance equations obtained in the following way: For a given entry site $e_k$ and pivot $\pi_k$, we {\em always disregard the violation of the hard-core constraints} when we compute the initial Boltzmann weight associated with $e_k$ and the final Boltzmann weight associated with various exits. In addition, we always include factors of $w$ to account for the presence or absence of dimers in the configuration associated with particular exits. 

If the worm move has started with an initial site $v_0$ that was defect free, then choosing the off-lattice exit corresponds to introducing a {\em pair of $\pm 1/2$ charges} at $e_0$ and $\pi_k$. Therefore, for worm moves that start in this way, the Boltzmann weight of an off-lattice exit must include a factor of $f_{1/2}^2$ relative to other exits to account for this pair of half-charge defects in the final configuration.  If, on the other hand, the worm move had started at an initial site $v_0$ that was entered via an off-lattice entrance ({\em i.e.}, initially hosted a half-charge defect), one way to do the book-keeping of factors of $f_{1/2}$ would be to incorporate a single factor of $f_{1/2}$ in the Boltzmann weight of a off-lattice exit relative to other exits at a subsequent pivot $\pi_k$, as well as at the initial pivot $\pi_0$.

However, this creates two cases and complicates the actual implementation. A little thought is enough to convince oneself that a completely equivalent procedure is to assign a factor of $f_{1/2}^2$ to {\em all} off-lattice exits at subsequent pivots $\pi_k$ {\em independent of whether $v_0$ initially had a half-charge defect on it or not}, {\em and to also assign the same factor of $f_{1/2}^2$} to the off-lattice exit that corresponds to aborting the worm and doing nothing at pivot $\pi_0$ in the case when the initial vertex $v_0$ initially had a half-charge defect on it and became the initial pivot $\pi_0$.

If we take this latter approach to the book-keeping, the book-keeping of factors of $f_{1/2}$ also has a unified description in all other eventualities.
When accounting for the Boltzmann weight of the other $z$ exit options, the Boltzmann weight of an  exit $x_k$ that initially hosts a half-charge defect (which will get healed if that exit is chosen) should be computed in this simpler-to-implement convention {\em without} taking into account the fact that this choice of exit will heal the half-charge defect, {\em i.e.}, {\em without any factor of $1/f_{1/2}$ associated with its weight}. This is because once this exit is chosen and the worm terminated, one would just have moved this half-charge defect to the intially defect-free vertex $v_0$ if one had started the worm with a defect-free $v_0$. If, on the other hand, the worm motion had started at a vertex $v_0$ that initially hosted a half-charge defect, then the Boltzmann weight of initial off-lattice exit at the initial pivot $\pi_0$ has already been assigned a factor of $f_{1/2}^2$ relative to the other initial exits at that stage itself, and again, this means that weight of the exit $x_k$ whose defect would be healed by choosing it {\em should have no factor of $1/f_{1/2}$ associated with it}.

More explicitly, these local detailed balance equations reduce to a set of $z+1$ equations that have the same structure as the directed loop equation sets studied in Ref.~\cite{Syljuasen_Sandvik_2002}:
	\begin{equation}
		\label{eq:detailed_balance}
		T^{\pi}_{ex}\omega^{\pi}_e = T^{\pi}_{xe}\omega^{\pi}_x,
	\end{equation}
	where $\omega^{\pi}_e$ and $\omega^{\pi}_x$ are the Boltzmann weights associated with the configurations corresponding to choices $e$ and $x$ for the worm head. The subscripts $e,x$ takes value from $0$ to $z$: the ``off-lattice" entrance or exit is indexed as $0$, while the other entrances or exits corresponding to the $z$ neighbors of $\pi$ are indexed from $1$ to $z$.
The weights that enter this equation are assigned using the rationale we have just outlined:
	\begin{eqnarray}
		\label{eq:weight}
		\omega^{\pi}_{x} = f_{1/2}^2 \delta_{x,0} + w^{n_d(\pi)}(1-\delta_{x,0})
	\end{eqnarray}
	where $n_d(\pi)=0,1$ is the number of dimers connected to the pivot $\pi$ if the corresponding exit has been chosen. The rationale for the factor $f_{1/2}^2$ in the first term of Eq.~\ref{eq:weight} has already been explained in detail above.

	It is straightforward to verify (see Sec.~\ref{appendix:detail_balance}) that any solution of the local detailed balance equations Eq.~\ref{eq:detailed_balance} ensures a valid rejection-free algorithm, {\em i.e.}, every worm constructed in accordance with the solution can be accepted with probability $1$. Note that the factors of $1/2$ used to abort the move right at the outset, as well as the fact that the next pivot is chosen ``myopically'', {\em i.e.}, without regard to the relative Boltzmann weights associated with the various choices, are also crucial for guaranteeing this property of the worm construction.
	
	Many choices solve such an under-determined set of equations, including the usual ``heat-bath'' (Gibbs sampler) solution $T^{\pi}_{ex}=\omega^{\pi}_x/\sum_{x'}\omega^{\pi}_{x'}$. In our study, we use a ``no-bounce'' solution whenever possible, or, if this is not possible due to a large imbalance in the value of the largest weight relative to the other weights, we use a ``one-bounce'' solution~\cite{Syljuasen_Sandvik_2002,Syljuasen_2003}. In addition, we test our results by comparing with those obtained using the heat-bath solution.

	Finally we note that the worm construction and its detailed balance conditions can be generalized to accommodate a nonzero density of unit-charge defects, with charge $\pm 1$. Furthermore, this method applies to any lattice with arbitrary coordination numbers. The proof of the detailed balance (see Appendix~\ref{appendix:detail_balance}) does not rely on the regularity of the underlying lattice, meaning this algorithm remains valid even for lattices with site-dependent coordination numbers.

	\begin{figure}[t]
		\centering
		\includegraphics[width=0.99\linewidth]{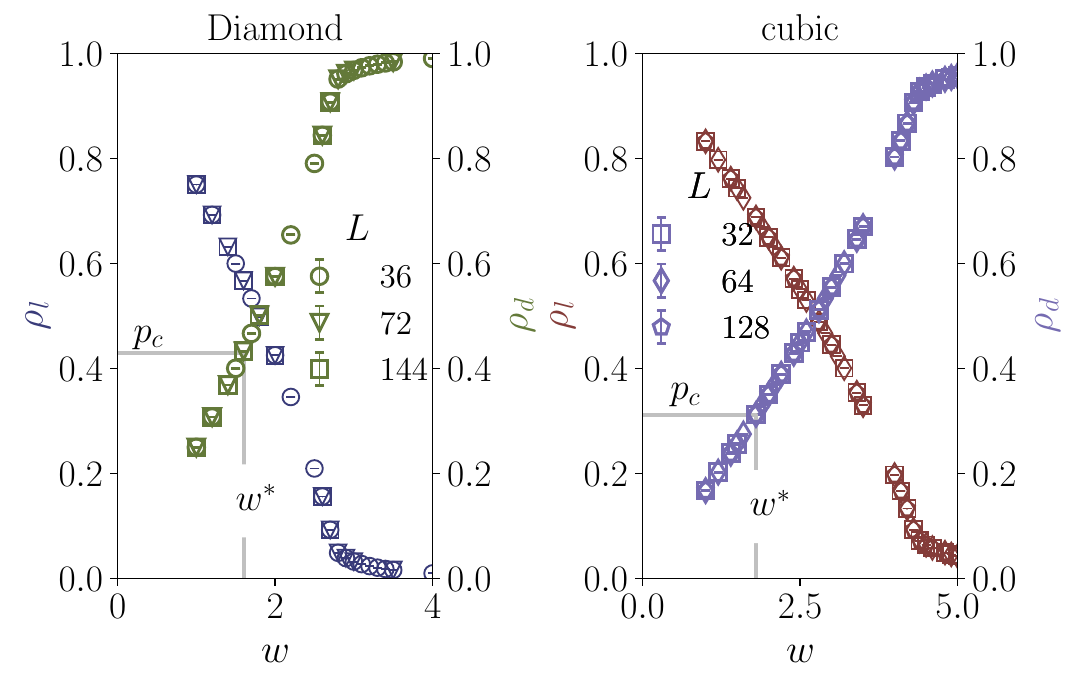}
		\caption{\label{fig:rho}Fraction of sites touched by nontrivial loops $\rho_l$ and the fraction of sites touched by trivial loops (dimers) $\rho_d$ plotted as a function of $w$ for both lattices. Note that these thermodynamic densities show no indication of a jump at any value of $w$.}
	\end{figure}

	\subsection{Measurements}
	\label{subsec:measurements}
	We have obtained data using large-scale Monte Carlo simulations performed on $L\times L\times L$ lattices with periodic boundary conditions, where $L$ is the number of unit cells along a principal direction of the lattice. In order to be able to perform a reliable finite-size scaling analysis, we have used a broad range of $L$, ranging from $18$ to $288$ ($16$ to $256$) for the diamond (cubic) lattice. 
	
	Apart from monitoring the thermodynamic densities $\rho_l$ and $\rho_d$, which respectively measure the fraction of sites that are touched by nontrivial loops and trivial loops (dimers), we have measured the distribution of loop sizes $P_l(s,L)$) as a function of $w$ in this range of $L$ on both lattices. Here, the size $s$ of a loop is defined as the total number of vertices it touches (this definition applies equally well to trivial loops, {\em i.e.}, dimers and open strings), and this distribution is obtained as the histogram of loop sizes in the loop ensemble being studied. $P_{l}(s,L)$ is related to $P_{\rm link}(s,L)$, 
	 the probability distribution of the size $s$ of the loop that passes through a randomly picked lattice site, via the correspondence
	 \begin{eqnarray}
	 P_{\rm link}(s,L) &= & s \cdot P_l(s,L) \; .
	 \end{eqnarray}
	 The factor of $s$ relating these two distributions has to do with the fact that a loop of size $s$ has an additional relative weight of  $s$ in the definition of $P_{\rm link}$. 
	 In order to make contact with scaling ideas~\cite{Jaubert_Haque_Moessner_2011,Nahum_Chalker_Serna_etal_2013_PRL} developed in previous studies of closely-related loop ensembles in three dimensions, we display in Sec.~\ref{subsec:fluxandloopsize} our results in the language of $P_{\rm link}(s,L)$ and not $P_l(s,L)$, although the latter is the quantity that is directly measured during our simulations. [We caution that Ref.~\cite{Kundu_Damle_2025} quoted results for $P_l(s,L)$ and not $P_{\rm link}(s,L)$ in the two-dimensional case studied earlier.]

	 We have also measured the flux distribution $P(\phi_x,\phi_y,\phi_z)$ for representative values of $w$ in its entire range. In the diamond lattice case, these fluxes are defined with respect to lattice planes whose normals are not perpendicular to each other (although we have labeled them with the subscripts $x$, $y$, $z$ to maintain uniformity in notation with respect to the cubic lattice case), but are instead aligned with the three principal axes of the diamond lattice along which we impose periodic boundary conditions. 
	
	 In  addition to keeping track of the size $s_{\rm max}$ of the largest loop in each dimer-loop configuration, we also measure  $S_{m}$ ($m=2,4$), the $m^{\rm th}$ moment of loop size, $\chi$, the loop size susceptibility, and the Binder cumulants $\mathcal{Q}_2$ and $\mathcal{R}$. These observables have the following definitions:
	\begin{eqnarray}
		\label{eq:chiandratios}
		S_m &=& \bigl\langle\sum_{j=1}^{n_l}s_j^m\bigr\rangle, \; \; \; \; \chi = S_2/N_s \nonumber \\
		\mathcal{Q}_2 &=& \bigl\langle\sum_{i\neq j}s_i^2 s_j^2\bigr\rangle/ 2\bigl\langle\sum_j s_j^2\bigr\rangle^2 = (S_2^2-S_4)/2S_2^2, \nonumber \\
		\mathcal{R} &=& \bigl\langle\sum_{j}s_j^4\bigr\rangle/\bigl\langle\sum_j s_j^2\bigr\rangle^2 = S_4/S_2^2,
	\end{eqnarray}
	where $N_s$ is the total number of sites, and $n_l$ is the total number of loops (including trivial loops) in a fully-packed configuration. 
	
	We also measure the positional correlations in equilibrium between a pair of half-vortices with charge $\pm 1/2$ introduced into an otherwise fully-packed equilibrium configuration by measuring the corresponding correlation function $C^{(1/2)}_v(\vec{r})$. $C^{(q)}(\vec{r})$,  the correlation function of two $\pm q$ test charges, is defined~\cite{Huse_Krauth_Moessner_Sondhi_2003} in general as the ratio of the partition functions with and without two test charges $\pm q$ separated by the displacement vector $
\vec{r}$:
	\begin{eqnarray}
		C^{(q)}_v(\vec{r}) = \frac{Z_q(\vec{r},w)}{Z(w)},
	\end{eqnarray}
	where $Z_q(\vec{r},w)$ represents the partition function of an otherwise fully-packed dimer-loop model with two defects of charge $\pm q$ separated by the displacement vector $\vec{r}$. By a slight abuse of notation, we use $C^{(q)}(r)$ to denote the $r=|\vec{r}|$ dependence of this correlation function $C^{(q)}(\vec{r})$ evaluated along the diagonal $x=y=z=r$ on both the diamond and the cubic lattice. Here, the coordinates $x$, $y$, $z$ are along the principal axes of the lattice in question (in the diamond lattice case, this corresponds to the orientations of three of the four body-diagonals of the up-pointing tetrahedron of the corresponding pyrochlore lattice whose vertices are defined by the centers of the diamond lattice bonds).
	
	From the arguments of Ref.~\cite{Alet_Ikhlef_Jacobsen_etal_2006}, it is clear that $C^{(1/2)}_v(\vec{r})$ can be computed by mapping it to the statistics of head-to-tail displacements in the half-vortex worm update used in our Monte Carlo simulation of the fully-packed case. Indeed, the measured histogram of the head-to-tail displacements is proportional to the corresponding correlation function $C_{v}^{(1/2)}(\vec{r})$ in our model.
	
	Using the analogous mapping to histograms of the head-to-tail displacements in the unit-vortex worm update, we also define a corresponding unit-vortex correlator $\widetilde{C}^{(1)}_v(\vec{r})$. However, we caution that the arguments of Ref.~\cite{Alet_Ikhlef_Jacobsen_etal_2006} do not really hold in detail in this case~\cite{Kundu_Damle_2025}. To see why, it is crucial to note the additional constraint inherent in our unit-vortex worm algorithm, which ensures that each unit vortex remains intact and does not fragment into two half-vortices before recombining at a later stage. 
	
	Due to this additional algorithmic constraint, the interpretation of $\widetilde{C}^{(1)}_v(\vec{r})$ defined in this way from the histogram of head-to-tail distances in the unit-vortex worm update is not straightforward. Indeed we will see that the behavior of $\widetilde{C}^{(1)}_v(\vec{r})$ is largely determined by the fraction of vertices that are visited by nontrivial loops, and its behavior admits an interesting interpretation in terms of a percolation transition in the large-scale geometry of regions of the lattice from which nontrivial loops are excluded. Further discussion of this point can be found in Sec.~\ref{subsec:corrfunc}.
	
	\begin{figure}[t]
		\centering
		\includegraphics[width=0.99\linewidth]{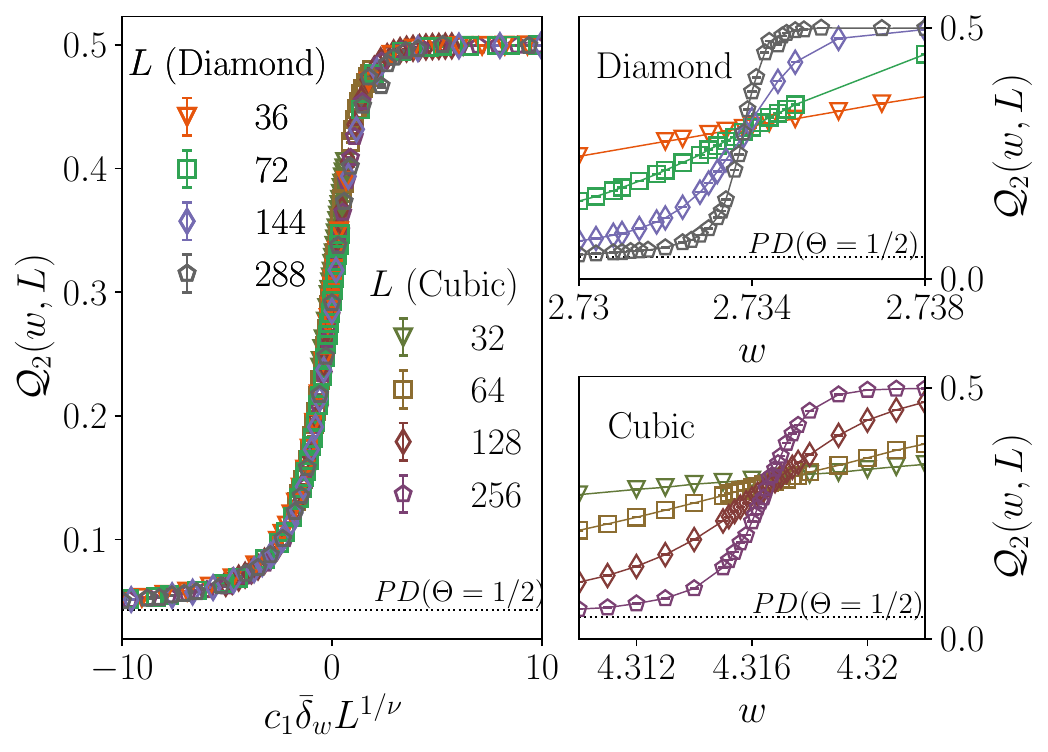}
		\caption{\label{fig:q2}Right panels: Binder ratio of loop sizes ($\mathcal{Q}_2$), as defined in Eq.~\ref{eq:chiandratios}, shows a clear crossing for different system sizes as a function of $w$, at a critical value $w_c=2.734(1)$ ($w_c=4.317(1)$) on the diamond (cubic) lattice. The black dotted horizontal line marks the theoretical prediction~\cite{Nahum_Chalker_Serna_etal_2013_PRL} for $\mathcal{Q}_2$ corresponding to a Poisson-Dirichlet (PD) form for the loop size distribution, with PD parameter $\Theta=1/2$ (see Eq.~\ref{eq:Q2andRprediction}). Left panel: Data for $w$ close to $w_c$ for different system sizes on both lattices collapse on to the scaling form defined in Eq.~\ref{eq:q2}. The scaling collapse displayed here employs the following parameter values: $w_c=2.7338$ ($w_c=4.3164$) for diamond (cubic) lattice, $\nu=0.63$ for both cases, and the lattice-dependent constant $c_1=1.0$ ($c_1=0.6$) for the diamond (cubic) lattice. We note that the choice $c_1=1$ for the diamond lattice is a convention used to define the scaling function $\mathcal{F}_{\mathcal{Q}}$ from the corresponding diamond lattice data collapse. }
	\end{figure}
	
	\begin{figure}[t]
		\centering
		\includegraphics[width=0.99\linewidth]{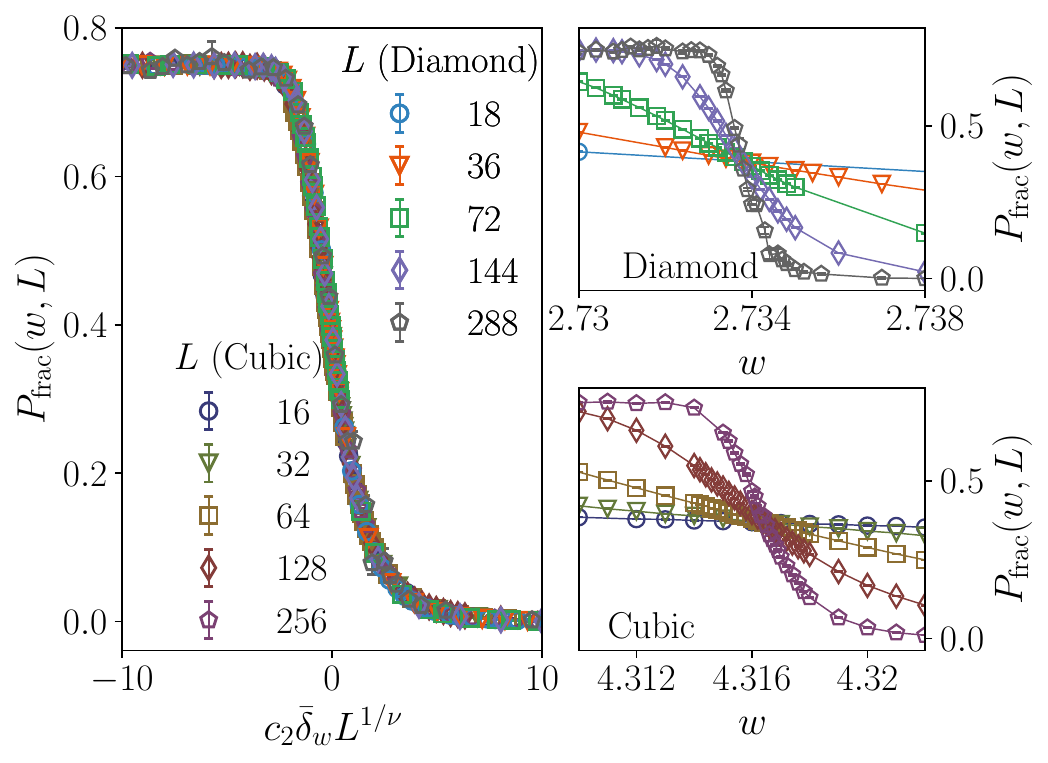}
		\caption{\label{fig:pfrac} Right panels: The probability of finding fractional fluxes, $P_{\rm frac}$ (defined in Sec~\ref{subsec:fluxandloopsize}) shows a clear crossing as a function of $w$ at a critical value $w_c=2.734(1)$ ($w=4.3165(10)$) on the  diamond (cubic) lattice. Left panel: Data for $w$ close to $w_c$ for different system sizes on both lattices collapses on to a single scaling form defined in Eq.~\ref{eq:q2}. The scaling collapse displayed here employs the following parameter values: $w_c=2.7338$ ($w_c=4.3166$) for diamond (cubic) lattice, $\nu=0.63$ for both lattices, and the lattice-dependent constant $c_2=1.0$ ($c_2=0.666$) for diamond (cubic) lattice. We note that the choice $c_2=1$ for the diamond lattice is a convention used to define the scaling function $\mathcal{F}_{P}$ from the corresponding diamond lattice data collapse.}
	\end{figure}
	
	%\newpage
	
	\section{Results}
	\label{sec:results}

	\subsection{Flux confinement-deconfinement transition}\label{subsec:fluxandloopsize}
	We find that the fraction of sites touched by trivial loops (dimers) $\rho_d$ increases monotonically as a function of $w$, without any sharp jump (note that the corresponding fraction of sites touched by nontrivial loops is $\rho_l \equiv 1-\rho_d$ by definition). Nevertheless, we see that there are two distinct thermodynamic phases separated by a continuous phase transition at $w_c=2.734(1)$ on the diamond lattice and $w_c=4.3165(10)$ on the cubic lattice. 
	
	This is evident from the $w$ dependence of the Binder cumulants $\mathcal{Q}_2(w)$ and the $w$ dependence of $P_{\rm frac}$, the probability that at least one of the three fluxes $\phi_x$, $\phi_y$ and $\phi_z$ is {\em not} an integer. As shown in Fig.~\ref{fig:q2} and \ref{fig:pfrac}, different curves of $\mathcal{Q}_2(w,L)$ and $P_{\rm frac}(w,L)$ for various system sizes intersect at $w_c\approx2.734(1)$ ($w_c=4.3165(10)$) on the diamond (cubic) lattice. Furthermore, near $w_c$, curves for different system sizes on both lattices collapse onto the scaling forms
	\begin{align}
		\label{eq:q2}
		\mathcal{Q}_2(w,L) &= \mathcal{F}_{\mathcal{Q}}(c_1\bar{\delta}_w L^{1/\nu})\nonumber,\\
		P_{\rm frac}(w,L) &= \mathcal{F}_{P}(c_2\bar{\delta}_w L^{1/\nu})
	\end{align}
	with a correlation length exponent of $\nu=0.63(1)$ on both lattices. Here $\bar{\delta}_w=(w-w_c)/w_c$ and $c_1,c_2$ are lattice-dependent scale factors which we set to $1$ for the diamond lattice by convention. 
	
	The value of $\mathcal{Q}_2$ jumps from $\mathcal{Q}_2^{<} \approx 0.043(1)$ for $w<w_c$ to $\mathcal{Q}_2^{>} \approx 0.50(1)$ for $w>w_c$. The observed value of $\mathcal{Q}_2^{<}$ is consistent with the fact that $w<w_c$ phase is dominated by the extended loops (of size $\propto L^3$), and the large loops have sizes that follow the Poisson-Dirichlet distribution~\cite{Nahum_Chalker_Serna_etal_2013_PRL}. On the other hand, the value of $\mathcal{Q}_2^{>}$ is consistent with the expectation for fully-packed configurations with short loops of length $\mathcal{O}(1)$. Thus, this transition is from an extended-loop phase to a short-loop phase. 
	
	In the large-size limit, the probability of half-integer fluxes, $P_{\rm frac}$, drops to zero for $w>w_c$, reflecting the fact that this transition should be thought of as a flux confinement-deconfinement transition: Half-integer fluxes proliferate in the $w<w_c$ phase, but are confined to integers in the $w>w_c$ phase. The restriction to integer fluxes in the $w>w_c$ phase is an emergent property of the phase, as opposed to a microscopic feature, in the sense that it does not directly follow from the microscopic constraints, which allow half-integer fluxes for all finite values of $w$; indeed, at a microscopic level, the requirement that fluxes be integer-valued only exists at $w = \infty$, for the limiting case of the fully-packed dimer model.
	
	\iffalse
	At $w = \infty$, the system corresponds to the classical dimer model, where fluxes are integer-valued by construction, and nontrivial loops are prohibited. In contrast, for finite values of $w$, nontrivial loops are permitted. Under our chosen normalization, this allows half-integer fluxes to appear, as the elementary defect charges at finite $w$ are half-integers. The smallest possible deviation from the zero-flux sector is thus $\pm \frac{1}{2}$, which occurs when a Dirac string carrying charges of $\pm \frac{1}{2}$ at its ends stretches across the open boundary of an infinite system or wraps around the system under periodic boundary conditions before recombining.
	
	For $w = \infty$, however, such configurations are not possible, as they would require the existence of nontrivial loops, which are forbidden in this regime. Consequently, at $w = \infty$, the elementary defect charges become $\pm 1$, and only integer flux sectors are accessible from the zero-flux sector. \fi
	
	The fact that the half-integer fluxes are excluded in the large-size limit when $w>w_c$ has interesting consequences for the dynamics of half-integer charge defects, which is reflected in the behavior of the half-vortex worms across the transition. This is explored further in Sec.~\ref{subsec:corrfunc}. 
	
	\subsection{Loop length distribution for $w<w_c$}
	\label{subsec:loopdistribution}
\begin{figure}[t]
		\centering
		\includegraphics[width=0.99\linewidth]{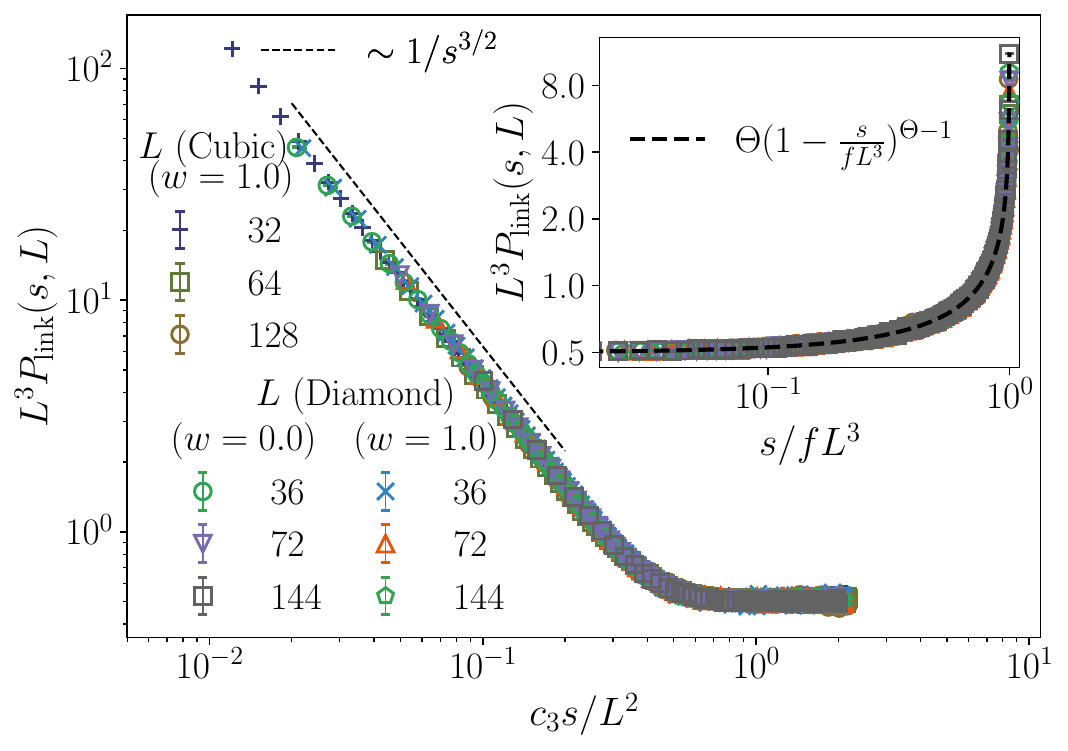}
		\caption{\label{fig:plinkwless} Main figure: The data for $P_{\rm link}(s,L)$ (as defined in Sec.~\ref{subsec:loopdistribution}) for different $w<w_c$ on both lattices ($w=0.0,1.0$ on the diamond lattice and $w=1.0$ on the cubic lattice) plotted in the range $s_0(w)\ll s\ll L^2$ where $s_0(w)$ is an ${\mathcal O}(1)$ size scale that depends on the lattice, and weakly on $w$. In this range of $s$, $P_{\rm link}(s)$ follow a power-law distribution $P_{\rm link}(s,L)\sim c_3/s^{3/2}$. $c_3$ is a lattice and $w$ dependent constant which is set to $1$ for the diamond lattice data at $w=0.0$ in our convention. Inset: In the limit $L^2\lesssim s \lesssim fL^3$, the data corresponding to $P_{\rm link}(s,L)$ for different $w$ on both lattices fit well to the Poisson-Dirichlet form described in Eq.~\ref{eq:pd}, with the Poisson-Dirichlet parameter $\Theta=1/2$. Here $f$ is a lattice and $w$ dependent fraction that corresponds to the fraction of sites that are touched by the largest loops on average; this fraction $f(w)$ can also be thought of as an intensive order parameter that characterizes the long-loop phase.}
	\end{figure}

	In this section, we characterize the long-loop phase by examining the loop-length distribution in more detail. Fig.~\ref{fig:plinkwless} displays the data for $P_{\rm link}(s,L)$, which represents the probability that a randomly selected site is part of a loop of length $s$. 
	
	We find that $P_{\rm link}(s,L)$ becomes largely independent of lattice-level details for $s$ larger than a microscopic size scale $s_0(w)$, typically of order a few lattice spacings for all $w$ in the long-loop phase. In this universal regime that is characteristic of the long-loop phase, $P_{\rm link}(s,L)$ exhibits two distinct behaviors depending on the size $s$ relative to the system size $L$: Loops of size $\xi(w)\ll s \ll L^2$ are largely unaffected by the finite size $L$ and exhibit characteristics similar to the Brownian motion. In this picture, $P_{\rm link}(s,L)$ corresponds to the return probability~\cite{Jaubert_Haque_Moessner_2011,Nahum_Chalker_Serna_etal_2013_PRL} of a random walker after $s$ steps; it is thus expected to follow a power-law distribution $P_{\rm link}(s,L)\propto s^{-d/2}$ in $d$ spatial dimensions. 
	
	Indeed, our numerical data in Fig.~\ref{fig:plinkwless} conforms to this expectation and is consistent with this picture. Further, we see that the data for $P_{\rm link}(s,L)$ for loop sizes $\xi(w)\ll s \ll L^2$ in the $w<w_c$ phase on both the lattices show scaling behavior:
	\begin{equation}
		P_{\rm link}(s,L) = \frac{1}{L^3}\mathcal{G}_{\rm link}(c_3 s/L^2)
		\label{eq:randomwalkscaling}
	\end{equation}
	with lattice-dependent constants $c_{3}$.  The quality of the fit of our numerical data in the main figure confirms this scaling. Further, we see that that $\mathcal{G}_{\rm link}(x)\sim 1/x^{3/2}$ for $x\ll1$, {\em i.e.}, it follows the power-law behavior expected by analogy to Brownian motion when $\xi(w)\ll s \ll L^2$. 
	
	However the tail of the distribution for $s \gg L^2$ follows a different limiting form: It crosses over to the Poisson-Dirichlet distribution for the {\em extended loops} of length $L^2\ll s \leq fL^3$, where $f$ corresponds to the fraction of sites that are touched by the largest loops on average; this fraction $f(w)$ can also be thought of as an intensive order parameter that characterizes the long-loop phase. 
	
	This is clear from the inset of Fig.~\ref{fig:plinkwless}, where our numerical data for $P_{\rm link}(s,L)$ in this { \em extended loop regime} is seen to fit well to the Poisson-Dirichlet form:~\cite{Nahum_Chalker_Serna_etal_2013_PRL}
	\begin{equation}
		\label{eq:pd}
		P_{\rm link}(s,L) = \frac{\Theta}{L^3}(1-\frac{s}{f L^3})^{\Theta-1},
	\end{equation}
	with the Poisson-Dirichlet parameter taking on the value $\Theta=1/2$.  
	
	\begin{figure}[t]
		\centering
		\includegraphics[width=0.99\linewidth]{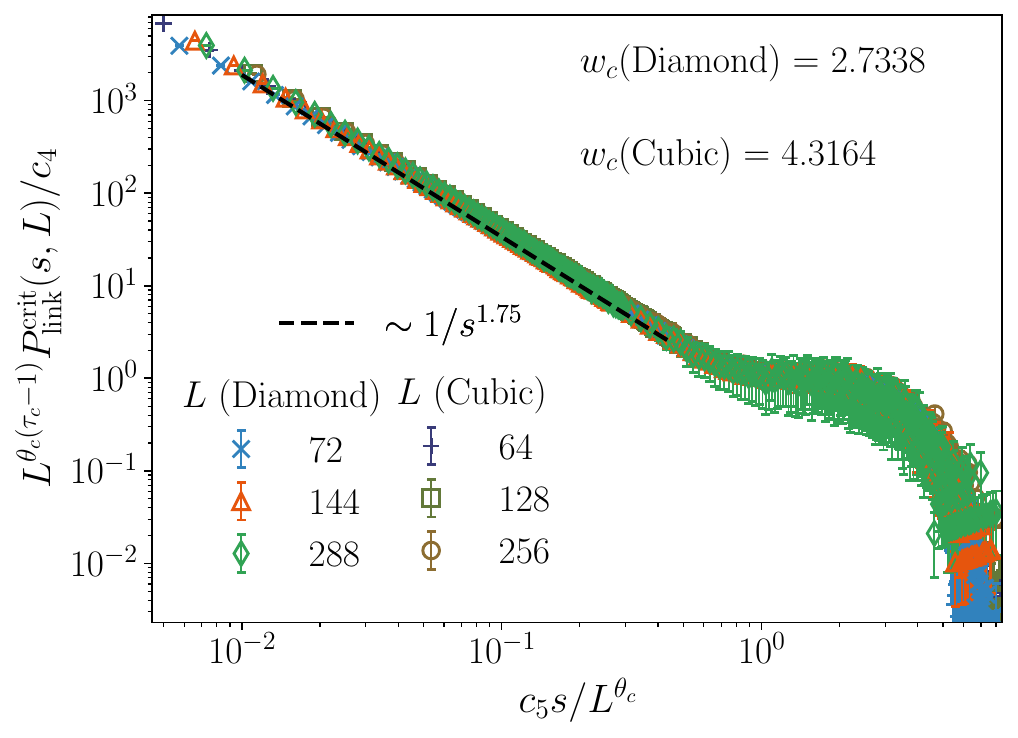}
		\caption{\label{fig:plcrit}The data for $P_{\rm link}^{\rm crit}(s,L)=sP_l^{\rm crit}(s,L)$ at the critical point $w_c=2.7338$ ($w_c=4.3164$) for diamond (cubic) lattice collapses on to a single scaling form described in Eq.~\ref{eq:plinkcrit} with universal exponent $\theta_c=1.72(3)$ and $\tau_c=2.75(2)$. $c_4$ and $c_5$ are lattice-dependent constants chosen with the convention that different system sizes of the diamond (cubic) lattice collapses on to the scaling form with $c_4=c_5=1$ ($c_4=2.0,c_5=1.6$). The scaling form reflects the fact that largest loop length at criticality is cutoff by a length scale $L^{\theta_c}$ and loops of length $s\ll L^{\theta_c}$ follow an universal power-law distribution $P_{l}^{\rm crit}(s,L)\sim 1/s^{\tau_c}$. }
	\end{figure}
	
	\begin{figure}[t]
		\centering
		\includegraphics[width=0.99\linewidth]{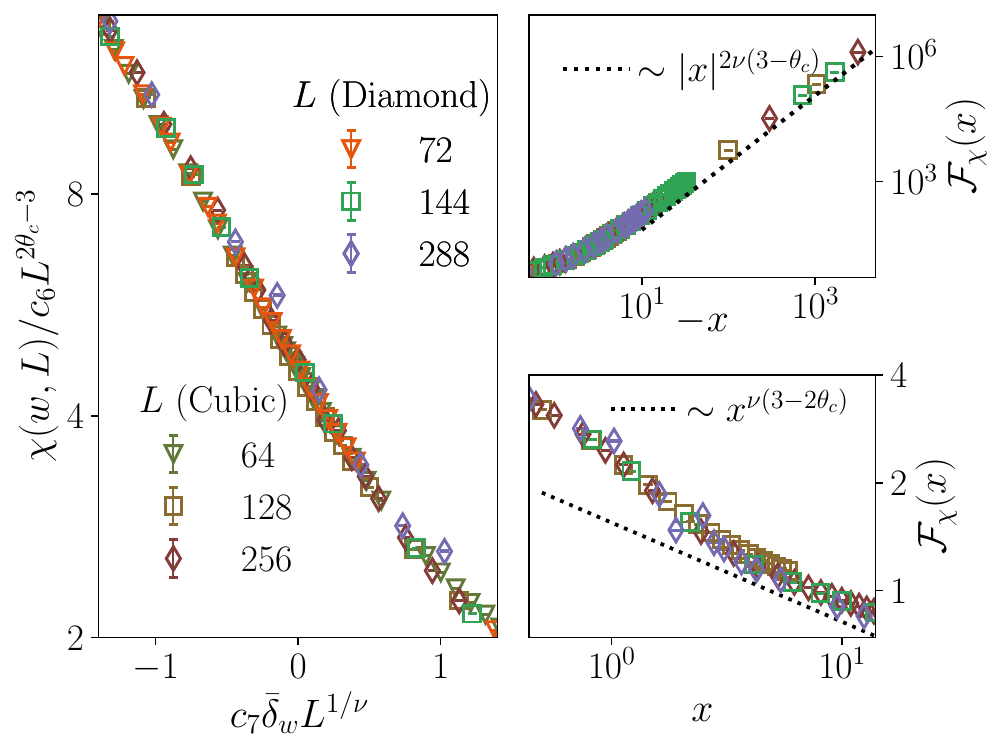}
		\caption{\label{fig:chi} Left panel: The loop size susceptibility on the diamond and the cubic lattices for different sizes in the vicinity of the critical point collapses on to the scaling form described in Eq.~\ref{eq:chi3d} with $\nu=0.63$ for both the lattices and $w_c=2.7338$ ($w_c=4.3164$) for diamond (cubic) lattice. The data displayed here are collapsed using $c_6=c_7=1$ ($c_6=0.42,c_7=0.611$) for the diamond (cubic) lattice. Right panels: Limiting power-law behavior of the scaling function $\mathcal{F}_{\chi}(x)$ for $w$ away from the critical point: For $x \ll 0$, $\mathcal{F}_{\chi}(x) \sim |x|^{2\nu(3-\theta_c)}$, while for $x \gg 0$, $\mathcal{F}_{\chi}(x) \sim x^{\nu(3-2\theta_c)}$, where $\theta_c = 1.72$ is extracted from the fits in Fig.~\ref{fig:plcrit} and $\nu = 0.63$ is obtained from the scaling collapses shown in Figs.~\ref{fig:q2} and \ref{fig:pfrac}. These limiting power-law behaviors reflect the fact that the loop size susceptibility scales as $L^3$ at $w\ll w_c$, being dominated by $\mathcal{O}(1)$ number of extended loops of size $s \sim L^3$. For $w\gg w_c$, the loop size susceptibility scales as $L^0$, since there are only short loops of $\mathcal{O}(1)$ length.}
	\end{figure}

	It is useful to emphasize a few points before we proceed further: First, the presence of a nonzero density of {\em extended loops} that wind around the system when periodic boundary conditions are imposed corresponds in the random walk picture to a nonzero probability for a random walker to escape to infinity in three dimensions. 
	
	This is in contrast to the long-loop phase of the corresponding dimer-loop model in two dimensions, where the largest loops have a size that scales as $s_{\rm max} \sim L^{1.75}$~\cite{Kundu_Damle_2025} and the fraction $f$ is zero. In the two-dimensional case, the power-law scaling of loop sizes for $\xi(w) \ll s \ll s_{\rm max}$ is also quite different from the expected scaling of the distribution of return times of a random walker. 
	
	Second,  our dimer-loop model maps at a microscopic level to the $O(1)$ loop model only at $w=0$. Nevertheless, the observed universal behavior of the loop size distribution throughout the $w<w_c$ long-loop phase phase does correspond to the expected behavior of the $O(1)$ loop soup in three dimensions~\cite{Nahum_Chalker_Serna_etal_2013_PRL}. In other words, the $w=0$ $O(1)$ loop model controls the large-scale behavior of the entire $w<w_c$ long-loop phase. 
	
	Third, the values of the Binder cumulants of the loop size distribution in the $w<w_c$ phase, as defined in Sec.~\ref{subsec:measurements}, are dominated by the loops which follow the Poisson-Dirichlet distribution. One indication of this is that $\mathcal{R}$ and $\mathcal{Q}_2$ both take on values consistent with the corresponding theoretical predictions for the Poisson-Dirichlet distribution in the long-loop phase:~\cite{Nahum_Chalker_Serna_etal_2013_PRL}
	\begin{align}
		\label{eq:Q2andRprediction}
		\mathcal{R} &= \frac{6(\Theta+1)}{(\Theta+3)(\Theta+2)},\nonumber\\
		\mathcal{Q}_2 &= \frac{\Theta (1+\Theta)}{2(2+\Theta)(3+\Theta)}
	\end{align}
This is displayed in Fig.~\ref{fig:q2} for $\mathcal{Q}_2$: The black-dotted horizontal line (labeled {\em PD}($\Theta=1/2$)) on each plot marks this theoretical prediction  (with $\Theta=1/2$) for $\mathcal{Q}_2$. We see that the data for different system sizes on both the lattices are consistent with this prediction for $w<w_c$. 

	\subsection{Critical loop length distribution}
	As one approaches $w_c$ in the long-loop phase, we find that the loop size distribution displays a crossover at a size scale $\xi(w)$, which increases rapidly with $w$, and diverges at the critical point. The fraction $f$ of sites visited by extended loops of size $s \sim L^3$ decreases correspondingly and goes to zero at $w_c$. For $ s_0 \ll s \ll \xi(w)$ in this regime (where $s_0$ is a microscopic size scale), the loop size distribution exhibits critical power-law behavior different from the random-walk behavior characteristic of the $O(1)$ loop model, although extended loops with $s \sim L^3$ continue to exist and the distribution displays this random-walk behavior beyond the size scale $\xi(w)$. 
	
	At criticality, we find that the loop size distribution on both the lattices obeys a universal scaling form:
	\begin{equation}
		\label{eq:plinkcrit}
		L^{\theta_c \tau_c-\theta_c} P_{\rm link}^{\rm crit} = c_4 \mathcal{G}_{\rm link}^{\rm crit}(c_5 s/L^{\theta_c}),
	\end{equation}
	with universal exponents $\theta_c=1.73(2)$ and $\tau_c=2.75(2)$ and lattice-dependent scale factors $c_{5,6}$. 	
	
	The universal scaling exponent $\theta_c$ encodes the fact that the maximum loop size at the critical point scales as $L^{\theta_c}$ on both the lattices, while $\tau_c$ determines the power-law form $1/s^{\tau_c}$ of the loop size distribution for $s_0 \ll s \ll L^{\theta_c}$. Equivalently, for  for $x\ll 1$,  the scaling function $\mathcal{G}_{\rm link}^{\rm crit}(x)$ takes on a power-law form $\mathcal{G}_{\rm link}^{\rm crit}(x)\sim 1/x^{\tau_c-1}$. The detailed evidence for these scaling properties is displayed in Fig.~\ref{fig:plcrit}, which should be compared with Fig.~\ref{fig:plinkwless} that displays the corresponding results in the long-loop phase.
	
	Note that the critical exponents $\theta_c$ and $\tau_c$ satisfy the relation $\theta_c\tau_c=d+\theta_c$ with $d=3$ within statistical error. As argued in Ref.~\cite{Kundu_Damle_2025}, the validity of such a scaling relation~\cite{Kondev_Henley_1995_PRL} has at its root the fact that the {\em number of distinct} large loops (of size scaling as $\mathcal{O}(L^{\theta_c})$) does not scale with the linear system size $L$ and remains of order one. It is also interesting to note that the scaling form Eq.~\ref{eq:randomwalkscaling}, which characterizes the Brownian part of loop size distribution ({\em i.e.}, the part that is modeled by the distribution of return times of a random walker) in the long-loop phase,  corresponds to the exponent values $\theta_{<} = 2$ and $\theta_{<} \tau_{<} = 5$, which also satisfy this scaling relation, although the rationale for its validity is not entirely clear in the random walk picture. 
	
	\begin{figure}[t]
		\centering
		\includegraphics[width=0.99\linewidth]{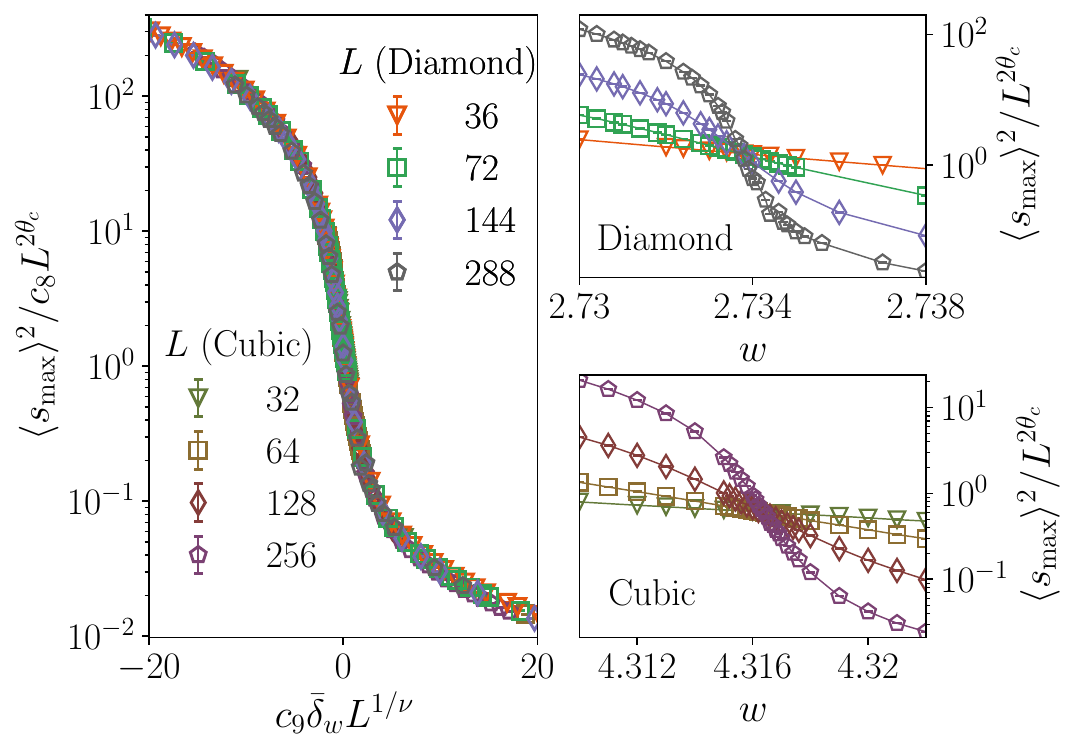}
		\caption{\label{fig:smaxsq} $s_{\rm max}$, the size of the largest loop, can be thought of as an order parameter for the long loop phase, since $s_{\rm max}$ scales as $L^3$ in the long loop phase. Right panel: $\langle s_{\rm max}\rangle ^2/L^{2\theta_c}$ plotted as a function of $w$ on the diamond and the cubic lattice, where $\theta_c$ is the exponent that controls the scaling of $s_{\rm max}$ at criticality via $s_{\rm max} \sim L^{\theta_c}$. For $\theta_c$ chosen equal to the value extracted from the scaling collapse of the critical loop size distribution in Fig.~\ref{fig:plcrit}, different curves corresponding to different system sizes cross at $w_c=2.7338$ ($w_c=4.3165$) on the diamond (cubic) lattice. Left panel: Data for $\langle s_{\rm max}\rangle^2/L^{2\theta_c}$ near the critical point on the two lattices follows the finite-size scaling form of Eq.~\ref{eq:smaxsq}. The scaling collapse shown here employs the following parameter values: $w_c=2.7338$ ($w_c=4.3165$) on the diamond (cubic) lattice and $\nu=0.63$ on both lattices, lattice-dependent scale factors $c_8=1,c_9=1$ ($c_8=0.4,c_9=0.6$) on the diamond (cubic) lattice, and $\theta_c$ obtained from the scaling collapse displayed in Fig.~\ref{fig:plcrit}. }
	\end{figure}
	
	\subsection{Susceptibility of loop length and the order parameter}
	\label{subsec:loopsusc}
	Loosely speaking, one may view $\langle s_{\rm max} \rangle$, the mean size of the largest loop in each configuration, as the ``order parameter'' for the long-loop phase. And stretching this analogy further, we may interpret  $\chi \equiv S_2/L^3$ as the ``loop size susceptibility''. We caution however that no actual broken symmetry is being probed by any of the local variables at our disposal.
	We find that the loop size susceptibility $\chi$, scales as $L^3$ in the $w<w_c$ long-loop phase, being dominated by a few extended loops of size $\mathcal{O}(L^3)$. In contrast, throughout the $w>w_c$ short-loop phase, $\chi$ is a $w$-dependent ${\mathcal {O}}(1)$ quantity. This stems from the fact that a typical loop configuration in the short-loop phase has $\mathcal{O}(L^3)$ short loops of $\mathcal{O}(1)$ length, with a loop size distribution that falls off exponentially beyond a characteristic $\mathcal{O}(1)$ size scale. For much the same reasons, $s_{\rm max}^2/L^3$ scales as $L^3$ in the long-loop phase, but goes to zero (as $1/L^3$) at large sizes in the short-loop phase. 
	
	In the neighborhood of the critical point, the loop size susceptibility is found to obey a universal crossover scaling function 
	\begin{equation}
		\label{eq:chi3d}
		\chi(w,L) = c_6L^{2\theta_c-3}\mathcal{F}_{\chi}(c_7 \bar{\delta}_wL^{1/\nu})
	\end{equation}
	with $\nu=0.63(1)$ and lattice-dependent scale factors $c_{6,7}$ (displayed in the left panel of Fig.~\ref{fig:chi}). Away from the critical point, $\mathcal{F}_{\chi}(x)\sim |x|^{2\nu(3-\theta_c)}$ for $x\ll 0$ and $\mathcal{F}_{\chi}(x)\sim x^{\nu (3-2\theta_c)}$ for $x\gg0$. The right panel of Fig.~\ref{fig:chi} shows the quality of the fits of our data corresponding to $\mathcal{F}_{\chi}(x)$ for $|x|\gg 0$.  Note that the conventional scaling form for a susceptibility in the vicinity of a critical point has a power-law prefactor $L^{2-\eta}$ instead of our prefactor of $L^{2\theta_c-3}$. With this in mind, one can equate the two forms of the prefactor to extract a value of $\eta$ from our fits by identifying $\eta$ with $5-2\theta_c$. However, we caution that the interpretation of this value for the anomalous dimension $\eta$ is not clear, since there is no local order parameter field associated with it.
	
	In an entirely analogous way, we find that $s_{\rm max}^2/L^3$ obeys the scaling form
		\begin{equation}
		\label{eq:smaxsq}
		s_{\rm max}^2/L^3 = c_8L^{2\theta_c-3}\mathcal{F^{'}}_{\chi}(c_9\bar{\delta}_w L^{1/\nu}),
	\end{equation}
	with lattice-dependent scale factors $c_8$ and $c_9$. This is shown in Fig.~\ref{fig:smaxsq}. Note that the scaling functions in the two cases have quite different limiting behavior for $x \gg 1$, since $s_{\rm max}^2/L^3$ vanishes at large sizes in the short-loop phase while $\chi$ goes to a nonzero constant.

	\begin{figure}[t]
		\centering
		\includegraphics[width=0.99\linewidth]{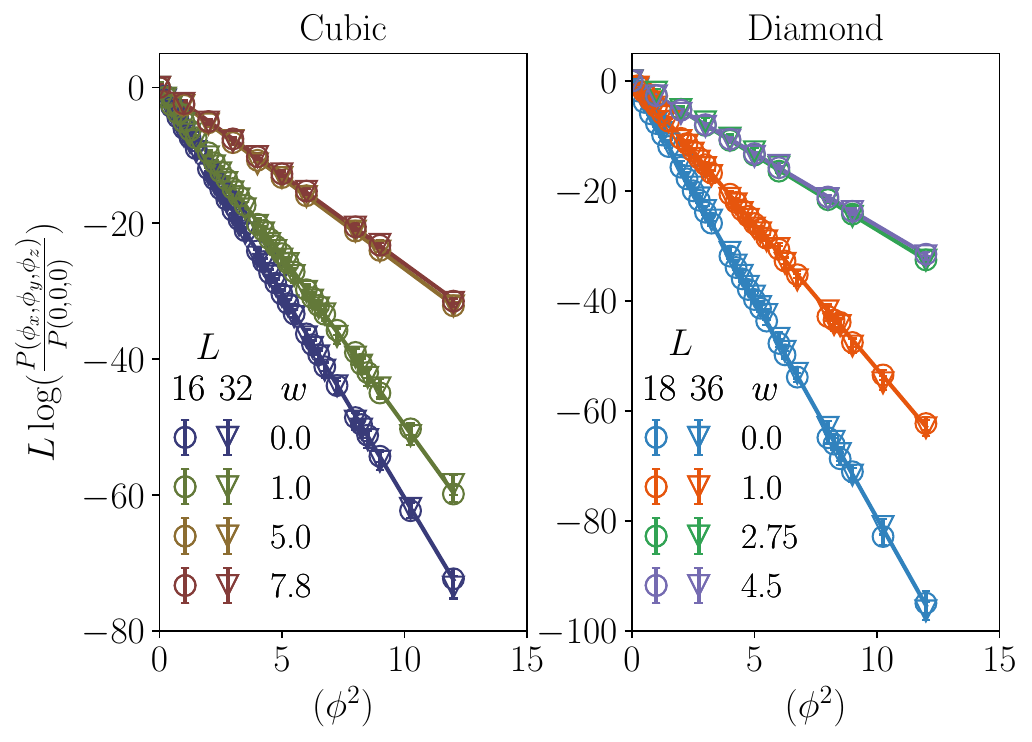}
		\caption{\label{fig:fluxdist} Data for $L \cdot \log(\frac{P(\phi_x,\phi_y,\phi_z)}{P(0,0,0)})$ plotted as a function of $\phi^2$ (as defined in Sec.~\ref{subsec:fluxdistribution}) on both lattices for different system sizes at two representative values of $w$ in each of the two phases. For $w<w_c$, data for all the flux configurations fits an $L$-independent straight line with a slope $-\kappa(w)$ that varies appreciably with $w$. For $w>w_c$ we see that fluxes belonging to the integer sectors, {\em i.e.}, with $(\phi_x,\phi_y,\phi_z)\in \{\phi\}_1$, follow this limiting Gaussian form, while flux vectors belonging to other sectors are exponentially suppressed in the large size limit. Also, in contrast to the substantial $w$ dependence of $\kappa$ in the long-loop phase, we find that $\kappa(w)$ has a relatively slight dependence on $w$ in the short loop phase. From linear fits of the data, we extract $\kappa(w)$ for different values of $w$ on both lattices: for the cubic lattice, $\kappa(0.0)=6.03(1)$, $\kappa(1.0)=4.98(3)$, $\kappa(5.0)=2.67(1)$, and $\kappa(7.8)=2.62(1)$; for the diamond lattice, $\kappa(0.0)=7.93(5)$, $\kappa(1.0)=5.26(5)$, $\kappa(2.75)=2.71(1)$, and $\kappa(4.5)=2.62(1)$.}
	\end{figure}

	\begin{figure}[t]
		\centering
		\includegraphics[width=0.99\linewidth]{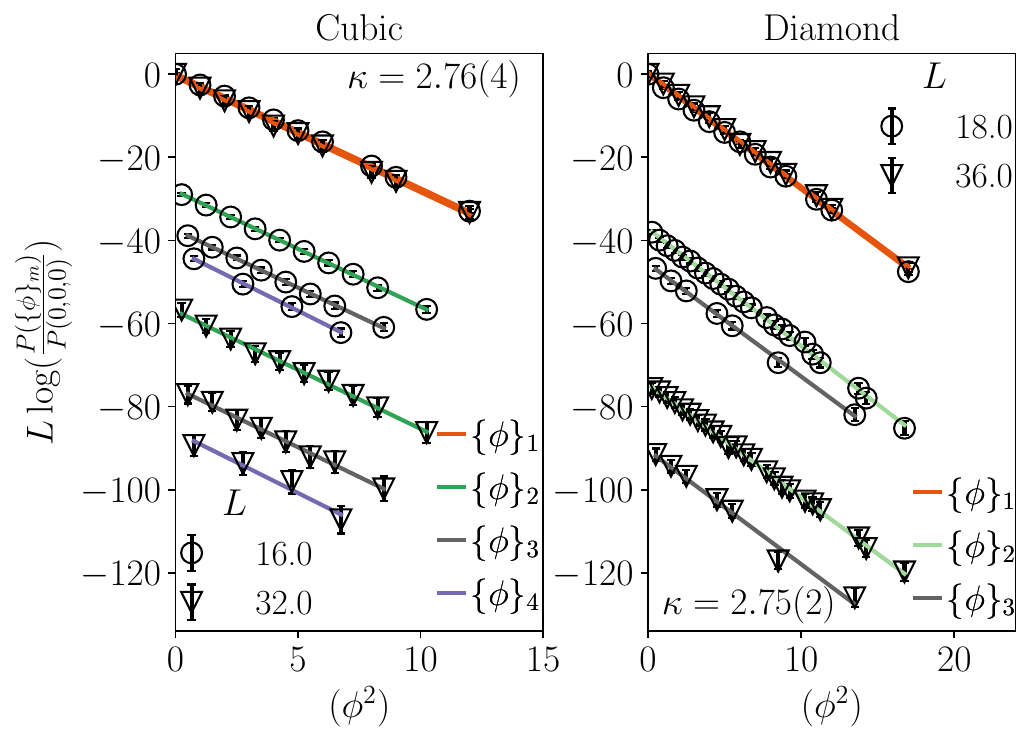}
		\caption{\label{fig:criticalfluxdistribution} Data at criticality for  the quantity $L \cdot \log(\frac{P(\phi_x,\phi_y,\phi_z)}{P(0,0,0)})$ corresponding to two different values of $L$ on each lattice, plotted as a function of $\phi^2$ in each case (as defined in Sec.~\ref{subsec:fluxdistribution}). We see that the cubic lattice dataset for two different $L$ seems to organize itself into seven different Gaussian behaviors at criticality. All these Gaussians have the same stiffness in the exponential, but different prefactors that multiply the exponential.  The diamond lattice data set for two different $L$ organizes itself in a similar way, but there are five different Gaussian behaviors seen. Again, they all have a common stiffness constant, but different prefactors. See Sec.~\ref{subsec:fluxdistribution} for a detailed discussion regarding the interpretation of these behaviors.}
	\end{figure}

	\begin{figure}[t]
		\centering
		\includegraphics[width=0.99\linewidth]{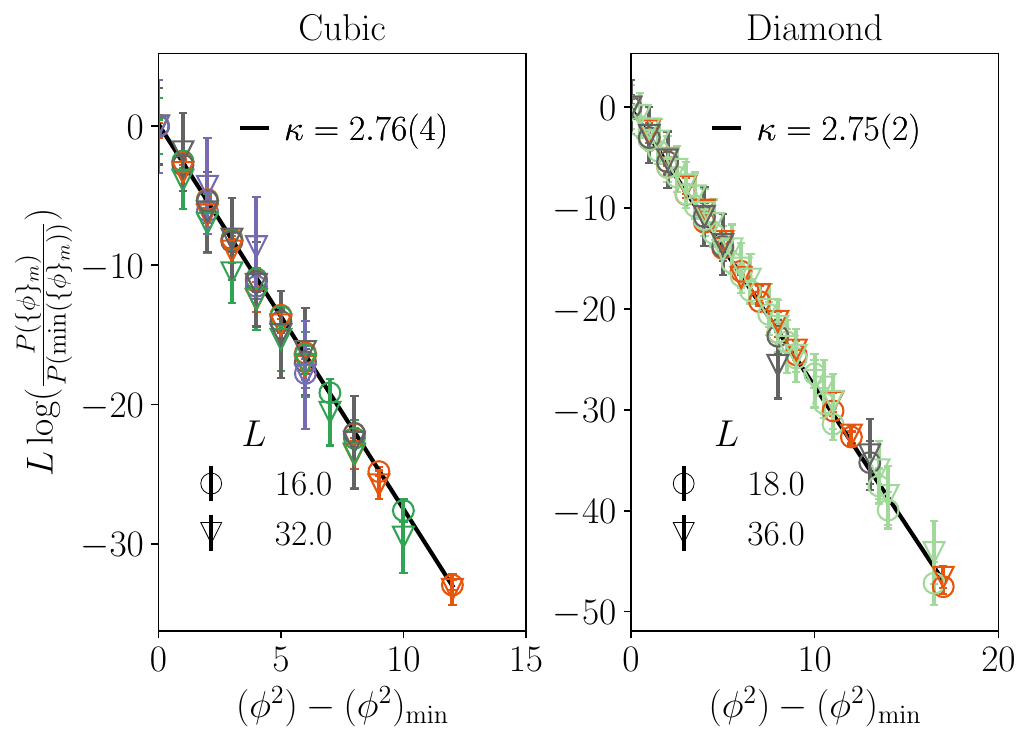}
		\caption{\label{fig:collapsedcriticalfluxdistribution} Data for $L \cdot \log(\frac{P(\{\phi\}_m)}{P({\rm min}\{\phi\}_m)})$ plotted as a function of
			$\phi^2 - (\phi^2)_{\rm min}$ separately for flux configurations belonging to each distinct subset 
			${\mathcal S}_m$ defined in Eq.~\ref{eq:diamondsectors} and Eq.~\ref{eq:cubicsectors} for the diamond and cubic lattice respectively. Here, $P(\{\phi\}_m)$ is the probability distribution $P(\phi_x,\phi_y,\phi_z)$ evaluated for flux sectors $(\phi_x, \phi_y,\phi_z) \in {\mathcal S}_m$, ${\rm min}\{\phi\}_m$ denotes a flux configuration 
			$(\phi_x, \phi_y, \phi_z) \in {\mathcal S}_m$ that has the smallest value  of 
			$\phi^2$ among all such configurations, and
			$(\phi^2)_{\rm min}$ is the corresponding minimum possible value of $(\phi^2)$ in that subset ${\mathcal S}_m$. }
	\end{figure}
	
	\subsection{Flux distribution}
	\label{subsec:fluxdistribution}
	As previously discussed in Sec.~\ref{subsec:Pmu}, the large-scale properties of both the fully-packed classical dimer model and the fully-packed O($1$) loop model on three-dimensional regular bipartite lattices can be described by the following effective action:
	\begin{equation}
		\label{eq:Seff3d}
		S_{\rm eff} = \frac{K}{2}\int d^3x (\mathbf{P}^2) \equiv \frac{K}{2}\int d^3x (\mathbf{\grad}\times\mathbf{A})^2 \; ,
	\end{equation}
	where $\mathbf{P}$ now represents the coarse-grained polarization field, and $\mathbf{A}$ is the corresponding coarse-grained vector potential. At a heuristic level, the field energy term $\mathbf{P}^2/2$ expresses an entropic preference for states with the largest possible number of ``nearby'' configurations, {\em i.e.}, configurations reached by local ring-exchange moves on flippable plaquettes.
	Since the dimer-loop model allows for a description in terms of a divergence-free polarization field for any nonzero $w$, we conjecture that the same effective action applies equally well for all  $w>0$, except possibly at the critical point $w_c$. 
	
	To test this picture, it is useful to work with a lattice-level effective action that accounts for the symmetries of the lattice~\cite{Huse_Krauth_Moessner_Sondhi_2003}, instead of the continuum theory above. To this end, we write
		\begin{equation}
		\label{eq:Slat3d}
		S_{\rm lat} = \frac{K}{2}\sum \mathbf{P}_l^2 \equiv \frac{K}{2}\sum (\mathbf{\Delta}\times\mathbf{A}_{\bar{l}})^2 \; ,
	\end{equation}
	where $ \mathbf{P}_l$ is a real-valued polarization vector on links of the lattice, and $\mathbf{A}_{\bar{l}}$ is a real-valued vector potential on links of the dual lattice. 
	
	The probability distribution $P(\phi_x,\phi_y,\phi_z)$ of a configuration characterized by net flux $(\phi_x,\phi_y,\phi_z)$ gives us a way of probing the form of this effective action and testing the validity of this picture. Indeed, if this picture is valid, one expects 
		\begin{equation}
		\label{eq:genflux}
		P(\phi_x,\phi_y,\phi_z) = C(L)\exp \left(-\frac{K}{2L}(\tilde{\phi}^2) \right) \; ,
	\end{equation}
where $C(L)$ is the normalization constant, and we caution that $\tilde{\phi}^2$, being short-hand for the quadratic form consistent with the structure of $S_{\rm lat}$, has different meanings on the cubic and diamond lattice:
\begin{eqnarray}
\tilde{\phi}^2 & \equiv &  \phi_x^2 + \phi_y^2 + \phi_z^2 \; {\rm (cubic \; lattice)} \nonumber \\
\tilde{\phi}^2 & \equiv &  \phi_x^2 + \phi_y^2 + \phi_z^2 +\phi_w^2 \; {\rm (diamond \; lattice),}
\end{eqnarray}
where $x$, $y$, $z$ denote the three principal directions of the cubic lattice, and, by a slight abuse of notation, we denote the orientations of the four different bonds of the diamond lattice in each unit cell (corresponding to the body diagonals of the up-pointing tetrahedron of the pyrochlore lattice formed by the centers of the diamond lattice bonds) by $x$, $y$, $z$, and $w$.

Note that the geometry of the diamond lattice, and the fact that we impose periodicity along three of these four directions, implies the constraint
\begin{eqnarray}
\sum_{\mu = x,y,z,w} \phi_{\mu} &=& 0 
\label{eq:constraintonfluxes}
\end{eqnarray} 
in every configuration of the dimer-loop model and one of these variables is redundant. Thus, in the diamond lattice case, we could equally-well have written the distribution as a function of just the three variables $\phi_x$, $\phi_y$,
		\begin{equation}
		\label{eq:diamondpflux}
		P(\phi_x,\phi_y,\phi_z) = C(L) \exp \left(-\frac{\kappa}{L}(\phi^2) \right) \;,
	\end{equation}
where $\phi^2 \equiv \tilde{\phi}^2/2 = \phi_x^2 + \phi_y^2+\phi_z^2+ \phi_x \phi_y + \phi_y \phi_z + \phi_z \phi_x$ and $\kappa = K$ in the diamond lattice case. In order to have a more unified notation in the subsequent analysis, we also define  $\phi^2 \equiv \tilde{\phi}^2$ and $\kappa = K/2$ on the cubic lattice, so that the flux distribution ansatz for the cubic lattice can also be rewritten in exactly the same form as Eq.~\ref{eq:diamondpflux}, {\em i.e.}, with $\kappa(\phi^2)/L$ in the exponential.

	With this in mind, we measure this distribution at representative points in both phases, as well as at the critical point.  As we will soon see, it is useful to introduce some additional notation and a classification of flux sectors in order to explain our observations regarding the flux distribution at the critical point and in both phases. Let $\{\phi\}$ represent the set of all possible flux configurations $(\phi_x,\phi_y,\phi_z)$. We split $\{\phi\}$ into three disjoint subsets depending on the value of $\phi^2$ on the diamond lattice.
	
	 These subsets are defined in the following way on the diamond lattice:
	\begin{eqnarray}
{\mathcal S}_1: \; \{\phi\}_1 &=&\{(\phi_x,\phi_y,\phi_z):\phi^2\in \mathbb{Z}\} \;, \nonumber \\
{\mathcal S}_2: \; \{\phi\}_2&=&\{(\phi_x,\phi_y,\phi_z):\phi^2\in \mathbb{Z}+1/4 \cup \mathbb{Z}+3/4\} \; , \nonumber \\
{\mathcal S}_3: \; \{\phi\}_3 &=& \{(\phi_x,\phi_y,\phi_z):\phi^2\in \mathbb{Z}+ 1/2\} \; .
		\label{eq:diamondsectors}
			\end{eqnarray}
	These three subsets correspond respectively to sectors with all three fluxes being integer-valued ($\{\phi\}_1$), sectors with one or two of the three fluxes being fractional ($\{\phi\}_2$), and sectors with all three fluxes being fractional ($\{\phi\}_3$). Note that the sector $\{\phi\}_2$ is a union of two distinct kinds of classes of configurations that are related by the symmetry of the diamond lattice. 
On the cubic lattice, these two classes are not related by any cubic lattice symmetry operation, and there are thus four distinct subsets we need to consider:
	\begin{eqnarray}
{\mathcal S}_1: \; \{\phi\}_1 &=&\{(\phi_x,\phi_y,\phi_z):\phi^2\in \mathbb{Z}\} \;, \nonumber \\
{\mathcal S}_2: \; \{\phi\}_2&=&\{(\phi_x,\phi_y,\phi_z):\phi^2\in \mathbb{Z}+1/4\} \; , \nonumber \\
{\mathcal S}_3: \; \{\phi\}_3 &=& \{(\phi_x,\phi_y,\phi_z):\phi^2\in \mathbb{Z}+ 1/2\} , \nonumber \\
{\mathcal S}_4: \; \{\phi\}_4&=&\{(\phi_x,\phi_y,\phi_z):\phi^2\in \mathbb{Z}+3/4\} \;\; .
		\label{eq:cubicsectors}
			\end{eqnarray}

In the long-loop phase, we find that the probability distribution of fluxes has a unified description, and follows  Eq.~\ref{eq:genflux} for all flux sectors on both lattices in the large size limit. However, in the short-loop phase, we find that all sectors other than those belonging to $\{\phi\}_1$ have a net weight that vanishes in the large size limit on both lattices, while flux configurations belonging to this subset of sectors continue to be described accurately by Eq.~\ref{eq:genflux}. 

This is clear from Fig.~\ref{fig:fluxdist}. Here we have plotted $L \cdot \log(\frac{P(\phi_x,\phi_y,\phi_z)}{P(0,0,0)})$ as a function of $\phi^2$ on both lattices for different system sizes and $w$ in both phases. For $w<w_c$, data for all the flux configurations fits into a $L$ independent straight line with a slope $-\kappa(w)$ that varies appreciably with $w$. For $w>w_c$ we see that fluxes belonging to the integer sectors, {\em i.e.}, with $(\phi_x,\phi_y,\phi_z)\in \{\phi\}_1$, follow this limiting Gaussian form, while flux vectors belonging to other sectors are exponentially suppressed in the large size limit. Also, in contrast to the substantial $w$ dependence of $\kappa$ in the long-loop phase, we find that $\kappa(w)$ has a relatively slight dependence on $w$ in the short loop phase. 

The critical point exhibits much more intricate behavior. We find that the probability distribution of fluxes  in all the sectors cannot be modeled by a single function of the form  Eq.~\ref{eq:genflux}. This is clear from Fig.~\ref{fig:criticalfluxdistribution}, which displays data for  $L \cdot \log(\frac{P(\phi_x,\phi_y,\phi_z)}{P(0,0,0)})$ for two different values of $L$ on each lattice, plotted as a function of $\phi^2$ in each case (as noted earlier, the definition of $\phi^2$ is slightly different on the two lattices). We see that the cubic lattice dataset for two different $L$ seems to organize itself into seven different Gaussian behaviors at criticality. All these Gaussians have the same stiffness in the exponential, but different prefactors that multiply the exponential.  The diamond lattice data set for two different $L$ organizes itself in a similar way, but there are five different Gaussian behaviors seen. Again, they all have a common stiffness constant, but different prefactors.

It is this crucial observation that motivates us to split the flux sectors into the different subsets ${\mathcal S}_m$ defined in Eq.~\ref{eq:diamondsectors} and Eq.~\ref{eq:cubicsectors} on the diamond and cubic lattices respectively. Analyzing the data separately in each of these subsets, we find that these different Gaussian behaviors seen in Fig.~\ref{fig:criticalfluxdistribution} arise from the following structure of $P(\phi_x, \phi_y, \phi_z)$:
		\begin{eqnarray}
		\label{eq:sectorialpflux}
		P(\phi_x,\phi_y,\phi_z) &=& C_m(L) \exp \left(-\frac{\kappa}{L}(\phi^2) \right) \nonumber \\
		 && \;\;\;\; \forall (\phi_x,\phi_y,\phi_z) \in {\mathcal S}_m \;,
	\end{eqnarray}
where ${\mathcal S}_m$ with for $m=1,2,3$ on the diamond lattice and $m=1,2,3,4$ on the cubic lattice are the subsets of flux sectors defined in Eq.~\ref{eq:diamondsectors} and Eq.~\ref{eq:cubicsectors} respectively, and the different prefactors $C_m(L)$ have different dependences on $L$ so that $C_m/C_{m'}$ also has nontrivial $L$ dependence when $m \neq m'$. 

It is this intricate  structure of $P(\phi_x,\phi_y,\phi_z)$ that is responsible for the fact that the cubic lattice dataset for two different $L$ appears to split into seven different Gaussian curves when plotted in the manner of Fig.~\ref{fig:criticalfluxdistribution}, while the corresponding diamond lattice data for two different $L$ splits into five different Gaussians. In general, if one plots the data for $N$ different sizes in the manner of Fig.~\ref{fig:criticalfluxdistribution} and there are $D$ different subsets of flux sectors with distinct prefactors $C_m(L)$, then one expects $ND-1$ different Gaussian curves in general, since $C_m/C_{m'}$ also has nontrivial $L$ dependence when $m \neq m'$, and forming the ratio $\frac{P(\phi_x,\phi_y,\phi_z)}{P(0,0,0)}$ only cancels out the prefactor for flux configurations belonging to ${\mathcal S}_1$ (to which the zero winding sector also belongs).

This structure of the critical flux distribution $P(\phi_x,\phi_y,\phi_z)$ is confirmed by the scaling collapse seen in Fig.~\ref{fig:collapsedcriticalfluxdistribution}, which plots $L \cdot \log(\frac{P(\{\phi\}_m)}{P({\rm min}\{\phi\}_m)})$ as a function of
 $\phi^2 - (\phi^2)_{\rm min}$ separately for flux configurations belonging to each distinct subset 
 ${\mathcal S}_m$. Here, $P(\{\phi\}_m)$ is the probability distribution $P(\phi_x,\phi_y,\phi_z)$ evaluated for flux sectors $(\phi_x, \phi_y,\phi_z) \in {\mathcal S}_m$, ${\rm min}\{\phi\}_m$ denotes a flux configuration 
  $(\phi_x, \phi_y, \phi_z) \in {\mathcal S}_m$ that has the smallest value  of 
  $\phi^2$ among all such configurations, and
   $(\phi^2)_{\rm min}$ is the corresponding minimum possible value of $(\phi^2)$ in that subset ${\mathcal S}_m$.

Since the prefactors $C_m(L)$ in Eq.~\ref{eq:sectorialpflux} cancel out when the ratios $\frac{P(\{\phi\}_m)}{P({\rm min}\{\phi\}_m)}$ are formed separately for each $m$, all data from all subsets ${\mathcal S}_m$ for various sizes $L$ is expected to collapse onto a single Gaussian when plotted as a function of $\phi^2 - (\phi^2)_{\rm min}$. This is exactly the scaling collapse established in Fig.~\ref{fig:collapsedcriticalfluxdistribution}, with the critical stiffness estimated to be $\kappa_c \approx 2.75(2)$ on the diamond lattice and $\kappa_c \approx 2.76(4)$ on the cubic lattice.

	\begin{figure}[t]
		\centering
		\includegraphics[width=0.85\linewidth]{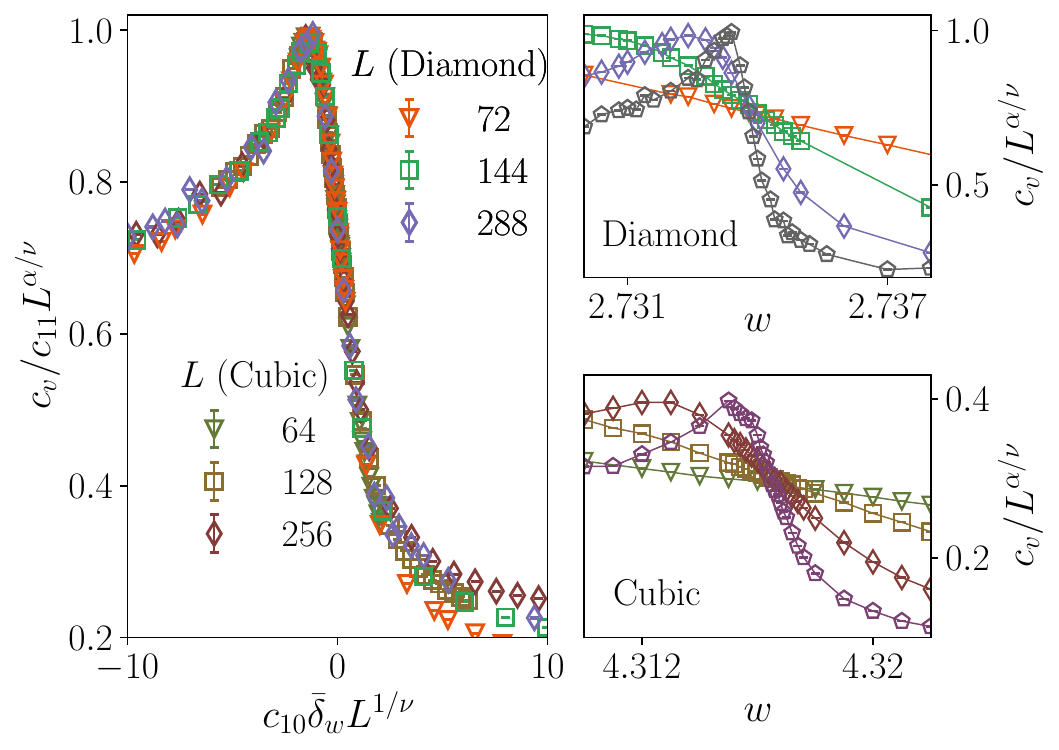}
		\caption[$c_v$ vs $w$.]{\label{fig:cv3d} Right panel: The data of the specific heat $c_v$ (defined in Eq.~\ref{eq:cv}) rescaled by $L^{\alpha/\nu}$ plotted as a function of $w$ for different system size on both lattices. With the choice of $\alpha/\nu=0.25(1)$ different curves corresponding to different system sizes cross at $w\approx 2.7338$ ($w=4.3166$) on the diamond (cubic) lattice and display a peak with same (size-independent) peak height for all system sizes. Left panel: Close to the critical point, data for $c_v$ exhibits scaling collapse and obeys the scaling form of Eq.~\ref{eq:cv3d}, with lattice-dependent parameters $c_{10}$ and $c_{11}$. The scaling collapse displayed here employs the following parameter values: $w_c=2.7338$ ($w_c=4.3165$) for the diamond (cubic) lattice and $\nu=0.63$ and $\alpha/\nu=0.25$ on both lattices. See Sec.~\ref{subsec:specificheat} for details}
	\end{figure}
	
	\subsection{Specific heat}
	\label{subsec:specificheat}
	The specific heat $c_v$ of the fully-packed classical dimer-loop model can be computed from the fluctuation of the number of trivial loops or dimers~\cite{Kundu_Damle_2025} via 
	\begin{eqnarray}
		\label{eq:cv}
		c_v = \frac{1}{L^3}\langle (n_d-\langle n_d\rangle)^2\rangle \; ,
	\end{eqnarray}
	where $n_d$ is the number of dimers or trivial loops in a fully-packed configuration of an $L \times L \times L$ sample. 
	
At a conventional second-order phase transition at which the usual finite-size scaling ideas are valid, we expect the specific heat of a sample with linear dimension $L$ to scale as $L^{\alpha/\nu}$ at criticality, where $\alpha$ is the specific heat exponent and $\nu$ is the correlation length exponent. If hyperscaling holds at a second-order critical point in $d=3$ dimensions, we also expect that
$\alpha = 2-3\nu$. 

If these ideas apply to our flux confinement-deconfinement transition, we thus expect
the specific heat to scale as $L^{2/\nu-3}$, where the correlation length exponent $\nu$ should take on a value consistent with our numerical estimate of $\nu \approx 0.63(1)$ obtained from the scaling of the probability $P_{\rm frac}$ of having non-integer fluxes and the Binder ratio of the loop sizes discussed in Sec.~\ref{subsec:fluxandloopsize},  as well as the scaling of the loop size susceptibility and loop size order parameter discussed in Sec.~\ref{subsec:loopsusc}. Therefore, we expect $\alpha/\nu \approx 0.17(5)$ if hyperscaling holds. 

However, as we see in Fig.~\ref{fig:cv3d}, we find that our data for $c_v$ admits a scaling collapse of the form	
	\begin{eqnarray}
		\label{eq:cv3d}
		c_v(w,L) = c_{11}L^{\alpha/\nu}\mathcal{F}_{c_v}(c_{10}\bar{\delta}_w L^{1/\nu}),
	\end{eqnarray}
	with lattice-dependent constants $c_{10}$ and $c_{11}$, and best-fit estimates of  $\alpha/\nu \approx 0.25(1)$ and $\nu \approx 0.63(1)$ for the exponents. Thus, our estimated value of $\alpha/\nu$ is inconsistent with the expected value of $\alpha/\nu$ predicted by hyperscaling using our own estimate of $\nu$.

Nevertheless, we emphasize that our value of $\nu$ is consistent with the known value of the correlation length exponent in the three-dimensional Ising universality class, although our estimate of $\alpha/\nu$ is of course inconsistent with the value of $\alpha/\nu$ for the three-dimensional Ising model since the three-dimensional Ising model obeys hyperscaling~\cite{Guillou_Justin_1980,Blote_Luijten_1995}. 
It would be interesting to further explore this apparent violation of hyperscaling as well as the fact that our value of $\nu$ does match that of the three-dimensional Ising model. We leave this as a challenge for future work.

	\subsection{Correlation between test vortices}\label{subsec:corrfunc}
 We have already seen in Sec.~\ref{sec:methodsandobservables3d} that $C^{(q)}_v(\vec{r})$,  the correlation function of two $\pm q$ test charges, is defined~\cite{Huse_Krauth_Moessner_Sondhi_2003} in general as the ratio of the partition functions with and without two test charges $\pm q$ separated by the displacement vector $
\vec{r}$. Our Monte-Carlo sampling procedure using the half-vortex worm algorithm provides an unambigious measurement of $C^{(1/2)}_v(\vec{r})$ via the measurement of the histogram of head-to-tail distances during the construction of the half-vortex worms. In contrast, as was already emphasized in Sec.~\ref{sec:methodsandobservables3d}, it is not straightforward to relate the corresponding histogram in the unit-vortex worm algorithm to the correlation function $C^{(1)}_v(\vec{r})$ of unit charges. 

Here we analyze our results for both histograms. From the half-vortex worm algorithm, we obtain a detailed characterization of the behavior of $C^{(1/2)}_v(\vec{r})$ in both phases and at the critical point. Moreover, we provide a simple interpretation of the $w$ dependence of the behavior of the corresponding unit-vortex histogram in terms of a percolation picture, which is consistent with the general idea that this measurement does {\em not} yield any direct information regarding the behavior of $C^{(1)}_v(\vec{r})$.
	
	\begin{figure}[t]
		\centering
		\includegraphics[width=0.99\linewidth]{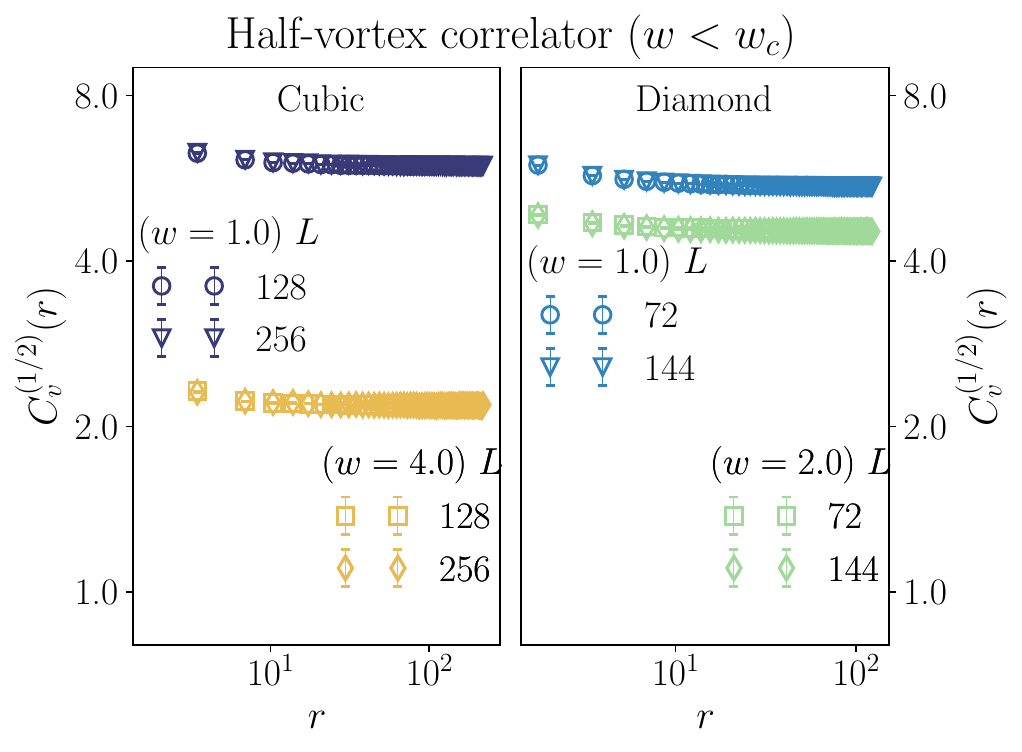}
		\caption{\label{fig:chalfwless} The half-vortex correlation function $C^{(1/2)}_v(r)$ for $w<w_c$ on the cubic and the diamond lattice saturates to a nonzero constant value at large distance.}
	\end{figure}
	
	\subsubsection{Correlation between a test-pair of  half-vortices}\label{subsubsec:corrfunchalfvortex}
	The half-vortex correlation function for $w<w_c$ on both lattices shows signature of deconfinement, analogous to the behavior of the monomer-antimonomer correlation function of the fully-packed dimer model on bipartite three-dimensional lattices~\cite{Huse_Krauth_Moessner_Sondhi_2003, Castelnovo_Moessner_Sondhi_review2012, Jaubert_Holdsworth_2009, Henley_Coulombphasereview2010}. In other words, $C^{(1/2)}_v(r)$, the correlation function for separation $\vec{r} = (x=r,y=r,z=r)$ (where $(x,y,z)$ are the components of $\vec{r}$ along the principal lattice axes for both the diamond and the cubic lattices) , decays to a {\em nonzero constant} at large $r$. The behavior in the short-loop phase is however quite different: $C_v^{(1/2)}(r)$ decays to zero faster than any power-law and is negligibly small already at $r \sim 10$ in lattice units.
	
\begin{figure}[t]
		\centering
		\includegraphics[width=0.99\linewidth]{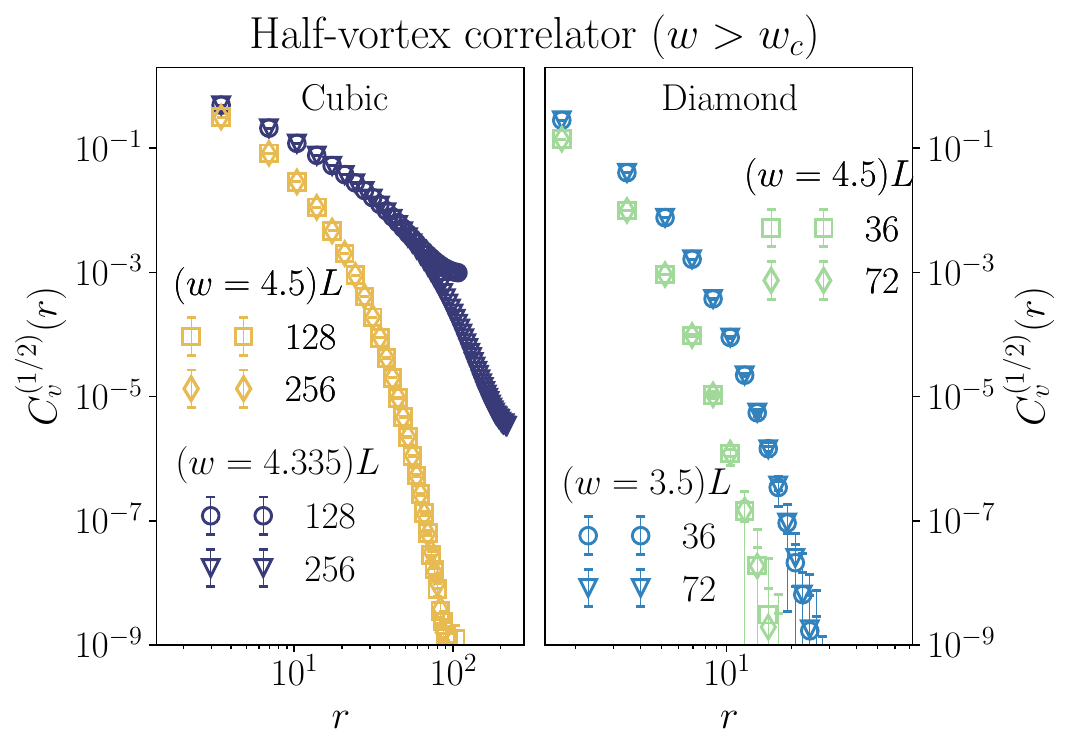}
		\caption{\label{fig:chalfwgreat} The half-vortex correlation function $C^{(1/2)}_v(r)$ for $w>w_c$ decays faster than any power on both the lattices.}
	\end{figure}

	The corresponding data is displayed in Fig.~\ref{fig:chalfwless} and \ref{fig:chalfwgreat}. The behavior in the long-loop phase is of course consistent with the general idea that the effective potential felt by one of these charges due to the other is an attractive Coulomb potential that falls off as $1/r$~\cite{Huse_Krauth_Moessner_Sondhi_2003, Castelnovo_Moessner_Sondhi_review2012, Jaubert_Holdsworth_2009, Henley_Coulombphasereview2010}. By the same token, the  contrasting behavior in the short-loop phase has a natural interpretation: For $w>w_c$, half-integer charges  are {\em confined} due to a strong attractive force between charges of opposite sign, corresponding to an effective  potential that grows linearly with separation $r$.

It is also useful to note that the our algorithmic prescription for measuring the correlation function $C_v^{(1/2)}(\vec{r})$  also provides a physical picture for the connection between the flux confinement transition at $w_c$ and these distinct behaviors of the correlation function $C_v^{(1/2)}(\vec{r})$ in the two phases separated by this transition: Consider a system in the zero flux sector. For this system to move to a half-integer flux sector, a worm has to wind around some periodic direction at least once, since the flux sector is a topological property of the system and no local changes of configuration can change the flux sector. However, since half-integer charges are confined, the probability of the worm being able to wind around a system of linear size $L$ falls off exponentially with $L$. In contrast, in the long-loop phase, half-integer charges are deconfined, allowing the system to freely access all flux sectors including those with fractional fluxes.

Of course, this argument is not complete, for it raises the question: How does the system transition to flux sectors with nonzero integer fluxes in the short-loop phase? After all, the probability distribution of integer flux sectors in the short-loop phase is a Gaussian with stiffness $\kappa/L$.  The answer has to do with the fact that unit-vortex worms, and the correlation function $\widetilde{C}_v^{(1)}(\vec{r})$ defined via the statistics of the head-to-tail displacement vectors $\vec{r}$ in the unit-vortex worm update, can and does have an entirely different behavior. This is what we turn to next.

	\subsubsection{Correlation between a test pair of full-vortices}
By keeping track of the histogram of the head-to-tail displacements of the worms constructed by the unit-vortex update, we have also computed the function $\widetilde{C}_v^{(1)}(r) \equiv \widetilde{C}_v^{(1)}(\vec{r} = (r,r,r))$.
In order to interpret our results for $\widetilde{C}_v^{(1)}(r)$, it is useful to recall that the unit-vortex worm lives on a {\em depleted lattice} consisting of vertices that {\em do not} have nontrivial loops passing through them. 

This feature of the unit-vortex worm construction is crucial, since it implies that the unit-vortex worm is trapped in finite clusters of the depleted lattice when the density $\rho_d$, defined as the fraction of sites touched by dimers, is smaller than the percolation threshold $p^*$ for the density $p$ of surviving sites on the parent cubic or diamond lattice. Likewise, for $\rho_d > p^*$,  we expect the unit-vortex worm to be able to explore the infinite connected cluster of the underlying depleted lattice in its geometrically percolated phase. Note that this is a purely kinematic restriction, quite independent of the nature of the effective potential between the unit-charged defects located on the head and the tail of the worm. It is for this reason that we cannot interpret our results for $\widetilde{C}_v^{(1)}(r)$ directly in terms of the effective potential between integer charges. [Note that this caveat applies equally well to the interpretation of the analogous results for $\widetilde{C}_v^{(1)}(r)$ in the two-dimensional case studied earlier in Ref.~\cite{Kundu_Damle_2025}.]

\begin{figure}[t]
		\centering
		\includegraphics[width=0.99\linewidth]{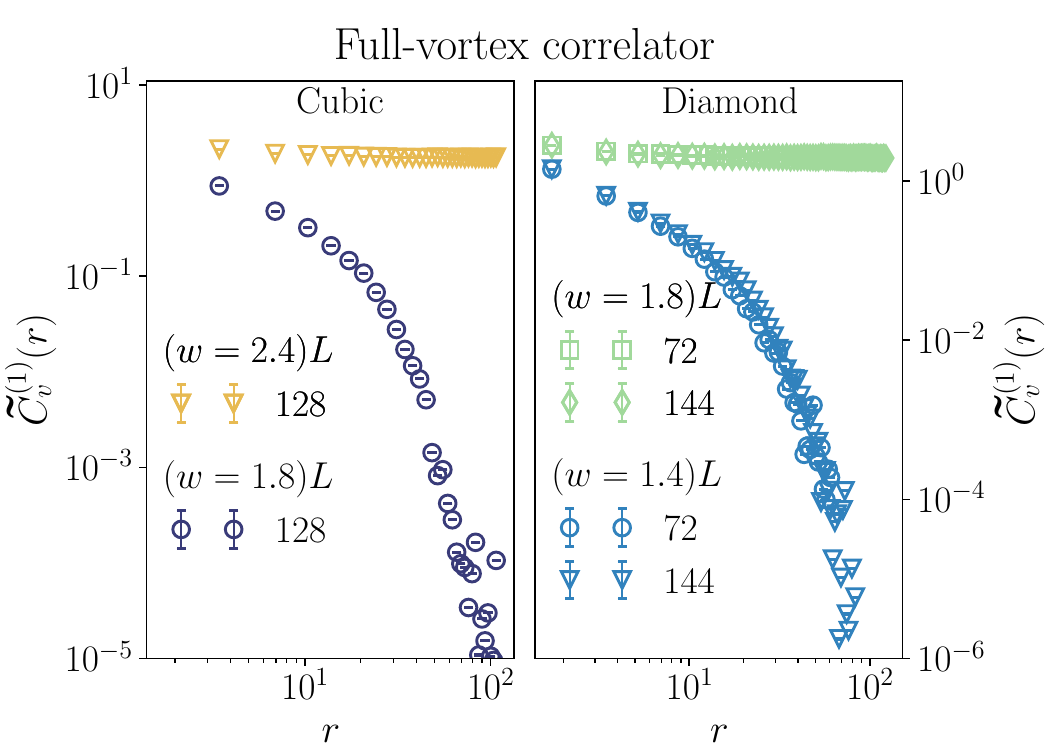}
		\caption{\label{fig:cfull} The correlator $\widetilde{C}^{(1)}_v(r)$, defined in terms of the histogram of the head-to-tail displacement of unit-vortex worms, is displayed on both lattices for two values of $w$ on either side of a threshold $w^{*} \neq w_c$ whose interpretation is provided in Sec.~\ref{subsec:corrfunc}. We see that it saturates to a nonzero constant at large separations $r$ for $w>w^{*}$, whereas it decays exponentially at large $r$ for $w<w^*$. We find that the threshold value separating these behaviors is $w^*\approx 1.6$ $(1.8)$ for the diamond (cubic) lattice. }
	\end{figure}

With all of this in mind, we now turn to the results for $\widetilde{C}_v^{(1)}(r)$. From our numerical data, we see that the large-$r$ asymptotics of $\widetilde{C}_v^{(1)}(r)$ changes qualitatively when $w$ is increased past  $w^*\approx 1.6$ ($w^*\approx 1.8$) for the diamond (cubic) lattice: For $w>w^*$, $\widetilde{C}_v^{(1)}(r)$ decays to a nonzero constant at large $r$. Whereas, for $w<w^*$, $\widetilde{C}_v^{(1)}(r)$ decays rapidly to zero, faster than any power law. These contrasting behaviors are displayed for representative values of $w$ on either side of $w^*$ in Fig.~\ref{fig:cfull}.

Crucially, this value of $w^{*}$ does not correspond to the critical coupling $w_c$ on either lattice. Moreover, the measured density of sites touched by dimers at $w^*$ is $\rho_d(w^*)=0.43(1)$ ($\rho_d(w^*)=0.31(1)$) on the diamond (cubic) lattice, which matches quite well with the classical site percolation threshold $p^{*}$~\cite{Ball_2014,Wang_Zhou_etal_2013,Frisch_Hammersley_1962,Sykes_Essam_1964}  of the diamond (cubic) lattice. This strongly suggests that the kinematic effect induced by the geometry of the depleted lattice is the dominant determinant of the large-$r$ asymptotics of $\widetilde{C}_v^{(1)}(r)$.

In other words, although the deconfined behavior of $\widetilde{C}_v^{(1)}(r)$ seen at large $w$ {\em does} imply that the actual unit-vortex correlator ${C}_v^{(1)}(r)$ will also be deconfined, the apparent ``confinement'' seen in the behavior of $\widetilde{C}_v^{(1)}(r)$ for $w< w^*$ is not a dynamical consequence of the effective entropic interaction between the charge $\pm 1$ defects located at the head and tail of the worm. Rather, it reflects the kinematics of motion on the depleted lattice. Further confirmation of this comes from the following: For the two-dimensional cases of the honeycomb and square lattice dimer-loop models studied in Ref.~\cite{Kundu_Damle_2025}, we have checked that this threshold behavior of the unit vortex correlator occurs at a value of $w^{*}$ that is again not the same as the location of the flux confinement-deconfinement transition (although, in those cases, it is much closer to $w_c$ than in the three-dimensional case studied here). Moreover, the value of $\rho_d(w^{*})$ again matches quite well with the known site-percolation thresholds on these lattices.

	\begin{figure}[t]
		\centering
		\includegraphics[width=0.99\linewidth]{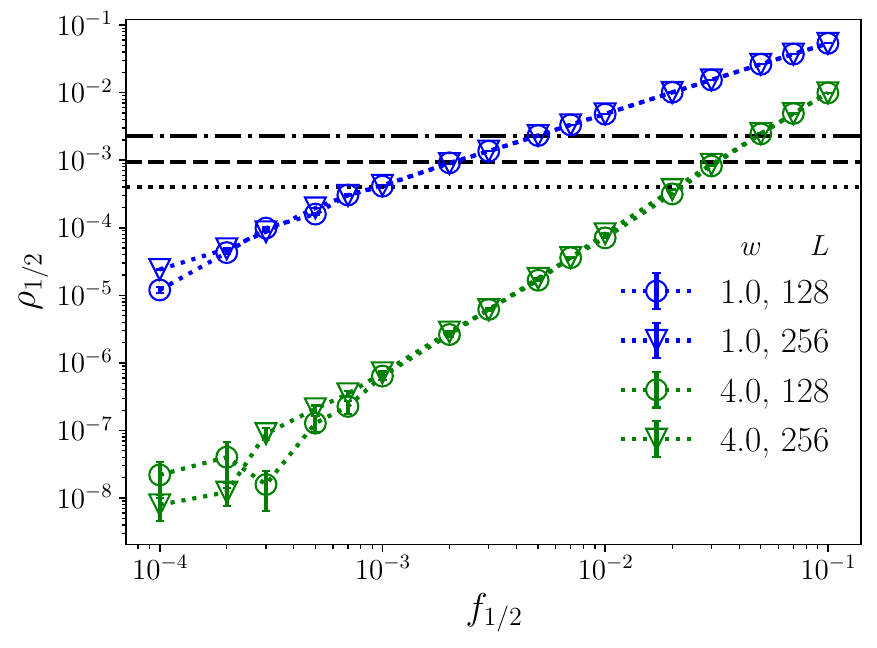}
		\caption{\label{fig:rhov} Density $\rho_{1/2}$ of half-integer vortices as a function of $f_{1/2}$ for $w=1.0$ and $w=4.0$. The horizontal lines mark different choices for the  common value of $\rho_{1/2}$ at which we compare the structure factor and two-point dipolar correlator at these two values of $w$ on either side of the flux confinement-deconfinement transition. The dotted line is at $\rho_{1/2}=0.0004$ that cuts the $w=1.0$ ($w=4.0$) curve at $f_{1/2}=0.001$ ($f_{1/2}=0.02$). Similarly, the dashed line is at $\rho_{1/2}=0.00095$ which cuts the $w=1.0$ ($w=4.0$) curve at $f_{1/2}=0.002$ ($f_{1/2}=0.035$) and the dash-dot line is at $\rho_{1/2}=0.0023$ which cuts the $w=1.0$ ($w=4.0$) curve at $f_{1/2}=0.005$ ($f_{1/2}=0.05$).}
	\end{figure}

	\subsection{Charge defects}\label{subsec:fhalfresults}
	As we have discussed in the previous section, test charges with charge $\pm 1/2$ are deconfined in the long-loop phase, but are confined in the short-loop phase. Naturally, this implies that a nonzero value for the fugacity $f_{1/2}$ of charge $\pm 1/2$ defects is a relevant perturbation of the fully-packed model in its long-loop phase. In addition, one may also be tempted to conclude that $f_{1/2}$ is an irrelevant perturbation of the fully-packed model in its short-loop phase. 
	
	However, this second conclusion is invalid for the following reason: Since two $+1/2$ charges at a small microscopic distance from each other are in effect equivalent to a charge $+1$ object as far as the long-distance physics is concerned, a nonzero value of $f_{1/2}$ is expected to induce an ${\mathcal O}(f_{1/2}^2)$  nonzero value of $f_{1}$ at the quadratic level. Therefore, if the fugacity $f_{1}$ for unit-charge defects is a relevant perturbation in the short-loop phase, then $f_{1/2}$ will also be marginally relevant in the sense that it will induce a relevant coupling at second order in $f_{1/2}$. 
	
	Further, since the physics of the short-loop phase is continuously connected to the physics of the fully-packed dimer model which constitutes its $w \rightarrow \infty$ limit, and since a nonzero value of $f_{1}$ is certainly a relevant perturbation of the fully-packed dimer model, we fully expect that $f_{1}$ is a relevant perturbation in the entire short-loop phase.
	Thus, we conclude that $f_{1/2}$ is a relevant perturbation (with a positive renormalization group eigenvalue) of the long-loop phase, while it is a marginally relevant perturbation (in the sense outlined above) in the short-loop phase. This argument is equally valid both in the two-dimensional cases studied in Ref.~\cite{Kundu_Damle_2025} and in the three-dimensional case studied here.

Next, we note that the ``bare'' value $\rho_{1/2}$ that is directly determined by the fugacity $f_{1/2}$ in our microscopic model just counts the number $n_{\rm open}$ of open strings (via $\rho_{1/2} \equiv 2 n_{\rm open}$). Thus $\rho_{1/2}$ is blind to the short-distance correlations between the termination points of such open strings. These correlations play a crucial role in determining the value of the coarse-grained charge density $\rho_{\rm eff}$ that corresponds to a given value of $\rho_{1/2}$.
To see this, consider a couple of simple examples: For instance, two such adjacent termination points correspond to a pair of equal and opposite charges, {\em i.e.}, a charge neutral object that does not contribute to $\rho_{\rm eff}$. On the other hand, two such termination points diagonally opposite each other on the same plaquette correspond to a pair of like charges, {\em i.e.}, a charge-$\pm 1$ defect which does contribute to $\rho_{\rm eff}$. 

If the scaling ideas described here are correct, one expects that a given microscopic value $\rho_{1/2}$ of charge $\pm 1/2$ defects would lead to a much larger value of $\rho_{\rm eff}$ in the long-loop phase compared to the short loop phase. Equivalently, the fugacity $y_v$ that controls the coarse-grained charge density $\rho_{\rm eff}$ in an effective long-wavelength description must have a very different dependence on the bare density $\rho_{1/2}$ in the two phases: $y_v$ is expected to increase much faster with $\rho_{1/2}$ in the long-loop phase compared to the short-loop phase, and therefore be much larger in the long-loop phase than in the short-loop phase for a given value of the bare density  $\rho_{1/2}$. 

To test this, we study the generalized dimer-loop model on the square lattice at two representative values of $w$ (one in each phase), with a small nonzero value of $f_{1/2}$, while keeping $f_1$, the fugacity of integer charge defects fixed at $f_{1} = 0$. In the microscopic model, this means we allow open strings with charge $\pm 1/2$ defects attached to their free ends, but do not allow any site to remain completely untouched by either a nontrivial loop or a trivial loop (dimer) or an open string.

In this generalized square lattice dimer-loop model, we measure the real space correlations of the bond occupation variable $n_{\mu}({\mathbf{r}})$ defined in Sec.~\ref{subsec:Pmu}, as well as the associated momentum space structure factor. From the real space correlators, we extract the ``dipolar correlator'' (more accurately, the dipolar part of the real-space correlation function) defined in Eq.~$45$ of Ref.~\cite{Desai_Pujari_Damle_2021}. This provides us a convenient way of probing the length scale beyond which the effect of a small nonzero $\rho_{\rm eff}$ (equivalently, a small nonzero fugacity $y_v$) become visible.

In a separate and more quantitative characterization, we also fit our data for the structure factor to the predictions of the coarse-grained theory~\cite{Desai_Pujari_Damle_2021} for the effect of a small nonzero fugacity $y_v$ on the dipolar pinch-point singularity of the structure factor of the Coulomb phase at wavevector $\mathbf{Q} = (\pi, \pi)$ on the square lattice. This allows us to extract best-fit values for $y_v$ in both phases for a fixed small value of $\rho_{1/2}$ and compare these values of $y_v$.

The equilibrium phase diagram of the square lattice dimer-loop model without charge defects has been characterized in detail in Ref.~\cite{Kundu_Damle_2025}, which established the presence of a flux confinement-deconfinement transition at $w_c=2.0$, which separates two Coulomb phases that are distinguished by the power-law columnar order present in the $w>w_c$ short-loop phase but absent in the $w<w_c$ long-loop 
phase. Here, we analyze the generalized dimer-loop model with nonzero $f_{1/2}$ for two representative values of $w$: $w=1$ in the long-loop phase and $w=4$ in the short-loop phase.
	\iffalse
	We define the dimer correlation function 
	\begin{equation*}
	C_{\mu\mu}(\mathbf{r}) = \langle n_{\mu}(\mathbf{r})n_{\mu}(\mathbf{0})\rangle-\langle n_{\mu}(\mathbf{r})\rangle\langle n_{\mu}(\mathbf{0})\rangle
	\end{equation*}
	For $w>2.0$, the dimer dimer correlation $C_{xx}(\mathbf{r})$ has the functional form 
	\begin{equation*}
		C_{xx}(\mathbf{r}) = (-1)^{r_x+r_y}f_d(\mathbf{r})+(-1)^{r_x}f_{\psi}(\mathbf{r})
	\end{equation*}
	where $f_{d}(r)$ is the dipolar correlation that has the asymptotic form $1/r^2$ and $f_{\psi}(r)$ is the columnar correlation. \fi

In Fig.~\ref{fig:rhov}, we plot the computed density of half-integer vortices, $\rho_{1/2}$, as a function of the fugacity $f_{1/2}$ for $w=1.0$ and $w=4.0$. For each of these two values of $w$, we  adjust $f_{1/2}$ for different $w$ to ensure that both systems have the same value of $\rho_{1/2}$, so that we may meaningully compare the behavior of the dipolar correlator and the structure factor for the same value of $\rho_{1/2}$. The horizontal lines in Fig.~\ref{fig:rhov} identify three such values of $\rho_{1/2}$ at which we have chosen to make such comparisons between the physics in the long loop phase (at $w=1.0$) and short-loop phase (at $w=4.0$). 

\begin{figure}[t]
		\centering
		\includegraphics[width=0.99\linewidth]{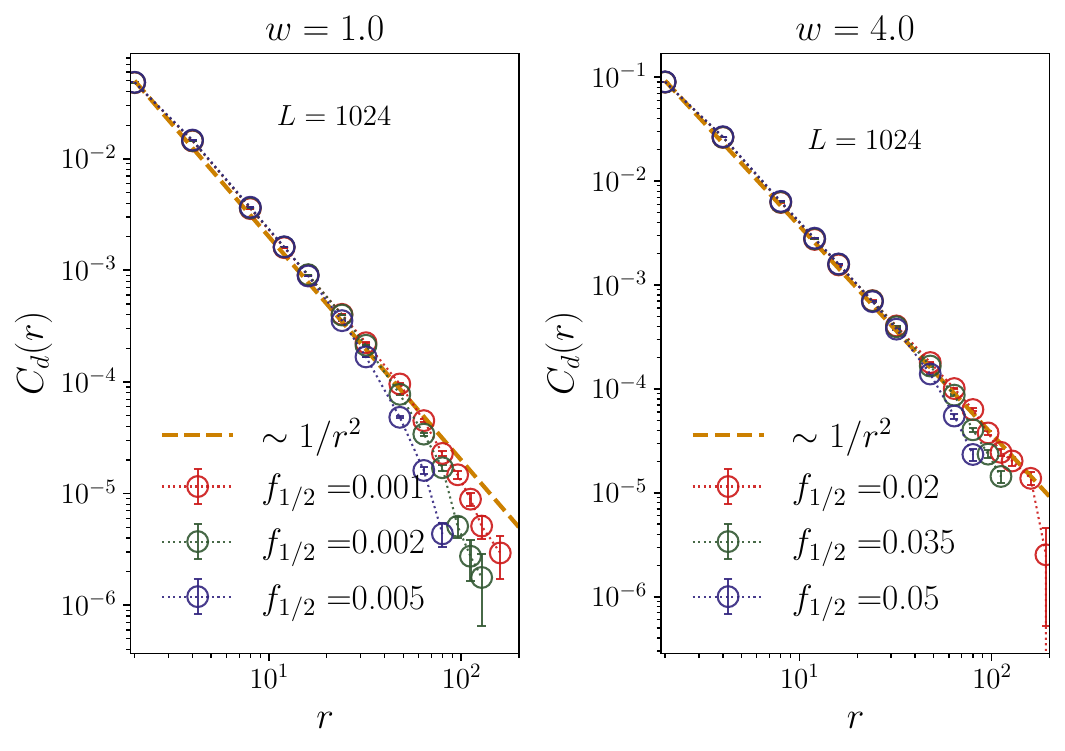}
		\caption{\label{fig:cdcrossover} The dipolar correlator $C_d(\mathbf{r})$ is defined (see Ref.~\cite{Desai_Pujari_Damle_2021}) as the linear combination $C_d(\mathbf{r})=(-1)^{r_x}(C_{xx}(\mathbf{r})-C_{xx}(\mathbf{r}+\hat{e}_y))$, where $C_{xx}(\mathbf{r}) = \langle n_x(\mathbf{r}) n_x(\mathbf{0})\rangle-\langle n_x(\mathbf{r})\rangle^2$. $C_d(\mathbf{r})$ is plotted as a function of $r$ at $w=1.0$ and $w=4.0$, for the values of $f_{1/2}(w)$ corresponding to the common values of the bare $\rho_{1/2}$ marked by horizontal lines in Fig.~\ref{fig:rhov}. Comparing these behaviors of  $C_d(\mathbf{r})$  at the two values of $w$ and the {\em common} value of $\rho_{1/2}$ marked by these horizontal lines in Fig.~\ref{fig:rhov}, we see that the crossover from the dipolar $1/r^2$ power-law behavior sets in at a much smaller length scale at $w=1.0$ compared to the corresponding crossover at $w=4.0$, although $\rho_{1/2}$ is the same in both cases. See Sec.~\ref{subsec:fhalfresults} for details.}
	\end{figure}

In Fig.~\ref{fig:cdcrossover}, we display our results for the dipolar correlator $C_d(r)$ for $w=1.0$ and $w=4.0$, with $f_{1/2}$ chosen  at each $w$ to yield one of the three values of $\rho_{1/2}$ marked in Fig.~\ref{fig:rhov}. From this data, we see quite clearly that the crossover from the $1/r^2$ dipolar power-law behavior to a faster decay as a function of $r$ occurs at a much smaller length scale for $w=1$ compared to $w=4$, although $f_{1/2}$ has been tuned to ensure that $\rho_{1/2}$ has the same value in the two cases. 
	
	\begin{figure}[t]
		\centering
		\includegraphics[width=0.99\linewidth]{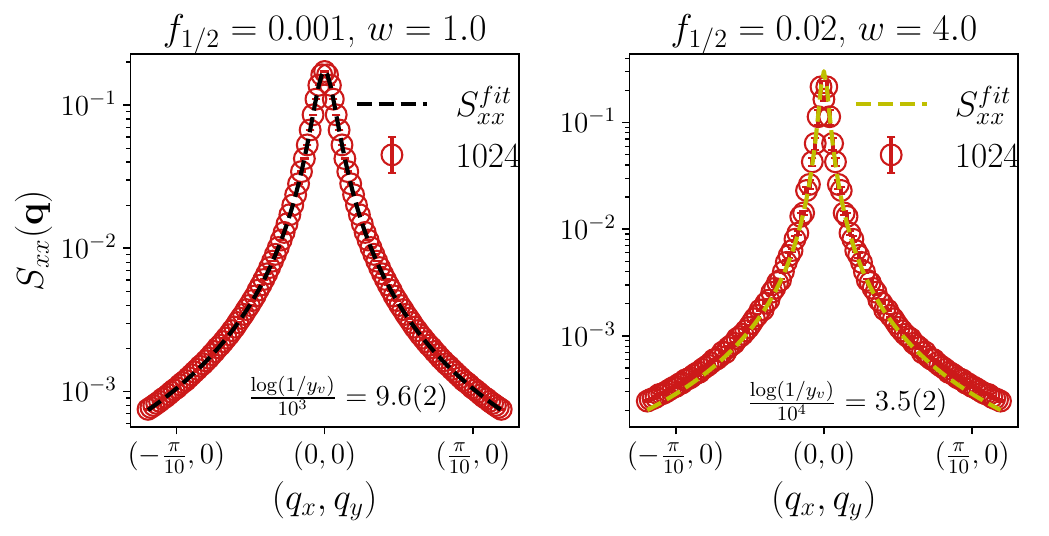}
		\caption{\label{fig:strfactor_pinchpoint} $S_{xx}(q_x)=\langle n_x(\mathbf{Q}+q_x\hat{k}_x)n_x(-\mathbf{Q}-q_x\hat{k}_x)\rangle$, the structure factor of link variables, plotted along the $q_x$ axes for $(w,f_{1/2})=(1.0,0.001)$ and $(4.0,0.02)$ in the vicinity of the pinch-point. Both curves are fit to the functional form in Eq.~\ref{eq:pinchpoint} to extract the phenomenological parameter $\log(1/y_v)$ that controls the coarse-grained charge density in the effective theory. Although both these ($w,f_{1/2}$) points correspond to the same bare value of $\rho_{1/2}$, the best-fit value of $\log(1/y_v)$ for $w=4.0,f_{1/2}=0.02$ is almost $4$ times larger than the best-fit value for $w=1.0,f_{1/2}=0.001$. See Sec.~\ref{subsec:fhalfresults} for details.}
	\end{figure}

	We also measure the structure factor
	\begin{equation*}
	S_{xx}(\mathbf{k}) = \langle n_x(\mathbf{k})n_x(-\mathbf{k}) \rangle
	\end{equation*}
	in the vicinity of the pinch-point wave vector $\mathbf{Q}=(\pi,\pi)$. As discussed in Ref.~\cite{Desai_Pujari_Damle_2021}, the presence of charge defects smears out the dipolar pinch-point singularity in $S_{xx}(\mathbf{Q})$. This effect is expected to be described by the following asymptotic form obtained from the effective theory~\cite{Desai_Pujari_Damle_2021}:
	\begin{eqnarray}
		\label{eq:pinchpoint}
		&\langle n_x(\mathbf{Q}+\mathbf{q})n_x(-\mathbf{Q}-\mathbf{q})\rangle \; \; \; \; \; \; \; \; \; \; \; \;\nonumber \\
		&\; \; \; \; \; \; \; \; \; \; \; \;= \frac{1}{\kappa}\frac{1+\frac{log(1/y_v)}{2\kappa}|f(q_y)|^2}{1+\frac{log(1/y_v)}{2\kappa}[|f(q_x)|^2+|f(q_y)|^2]} \nonumber\\ \nonumber \\
		&\langle n_y(\mathbf{Q}+\mathbf{q})n_y(-\mathbf{Q}-\mathbf{q})\rangle \; \; \; \; \; \; \; \; \; \; \; \;\nonumber \\
		&\; \; \; \; \; \; \; \; \; \; \; \;= \frac{1}{\kappa}\frac{1+\frac{log(1/y_v)}{2\kappa}|f(q_x)|^2}{1+\frac{log(1/y_v)}{2\kappa}[|f(q_x)|^2+|f(q_y)|^2]},
	\end{eqnarray}
	where $f(x)=1-e^{ix}$, $\kappa$ is the Gaussian stiffness parameter and $y_v$ is the effective fugacity parameter. From these functional forms, it is evident that the pinch-point structure is restored in the limit 
	$\log(1/y_v) \to \infty$, while it is entirely destroyed when $\log(1/y_v) = 0$. Consequently, the quantity $\log(1/y_v)$ provides a natural measure of the stability or resistance to the destruction of the pinch-point structure. 
	
	In Fig.~\ref{fig:strfactor_pinchpoint} we display this structure factor
	along the $k_x$-axis near $\mathbf{Q}$ for the two parameter sets: $(w, f_{1/2}) = (1.0, 0.001)$ and $(4.0, 0.02)$; as already noted, the measured density $\rho_{1/2}$ is equal at these two points. This data is fitted to the functional form given in Eq.~\ref{eq:pinchpoint}, which allows us to extract the fitting parameter $\log(1/y_v)$.

From this analysis, we find that the best-fit value of $\log(1/y_v)$ for $w=4.0,f_{1/2}=0.02$ is approximately four times larger than that of $w=1.0,f_{1/2}=0.001$ although the measured value of $\rho_{1/2}$ is identical at these two points. This is exactly what we expect if $y_v$ increases much more rapidly with $\rho_{1/2}$ in the long-loop phase compared to the short-loop phase.

Thus, our analysis of the real space dipolar correlator as well as the structure factor in the vicinity of the pinch-point singularity yields results that support the scaling picture discussed at the outset in this section. This is our basic conclusion. To go beyond this, one would need much more extensive numerics and we leave a more detailed analysis of this sort to future work.

	\section{Discussion}
	\label{sec:Discussion}
	This study investigates the statistical physics of the fully-packed dimer-loop model on three-dimensional bipartite lattices and uncovers a continuous phase transition between two dipolar Coulomb phases, tuned by the dimer fugacity $w$. These two Coulomb phases possess distinct topological characterizations: for periodic samples, half-integer fluxes (along each periodic direction) of the polarization field proliferate in the $w<w_c$ phase, while they are confined to integer values in the $w>w_c$ phase.
	Equivalently, and independent of boundary conditions, half-integer test charges $q=\pm 1/2$ are confined for $w>w_c$ but become deconfined in the small-$w$ phase.
	Although we have loosely used $\langle s_{\rm max}\rangle$ as an ``order parameter", we emphasize that no actual broken symmetry is detected by any local variable at our disposal. Whether a spontaneous symmetry breaking of some local (Ising-like) degrees of freedom exists remains an open question, which we leave for future exploration.
	
	Critical scaling reveals a striking combination of features. The correlation-length exponent $\nu$ is found to be very close to the three-dimensional Ising value, reminiscent of similar behavior in two dimensions~\cite{Kundu_Damle_2025}.
	However, the specific-heat exponent $\alpha$ differs significantly from that of the 3D Ising model, signaling an apparent violation of hyperscaling. Whether this represents a genuine breakdown of hyperscaling or a subtle crossover effect remains unresolved.
	The near-Ising value of $\nu$ in the absence of full Ising universality is particularly interesting and calls for deeper theoretical understanding.
	
	Through a spin-mapping construction, detailed in the context of the related two-dimensional dimer-loop model in Ref.~\cite{Kundu_Damle_2025}, the physics of the diamond and cubic-lattice dimer-loop models can be related to anisotropic spin-$S=1$ antiferromagnets on pyrochlore and corner-sharing octahedral lattices at certain magnetization plateaus.
	However, realizing a corresponding spin model with global Ising anisotropy on 3D pyrochlore or corner-sharing octahedral lattices is challenging because of the intrinsic crystal-field anisotropy, which aligns local quantization axes along the tetrahedral (or octahedral) directions~\cite{Harris_etal_1997,Siddharthan_Shastry_etal_1999,Fennel_etalScience2009,Castelnovo_Moessner_Sondhi_review2012,Bramwell_Harris_review2020}.
	
	Nevertheless, non-Kramers pyrochlore magnets with small crystal-field splittings could, in principle, exhibit similar physics if ``internal" effective magnetic fields arising from pseudospin doublet splittings mimic the required anisotropy. Such materials may thus provide a natural platform to explore these distinct topological phases and their unconventional critical behavior.
	
	\section*{Acknowledgments} 
	We thank F.~Alet, S.~Bhattacharjee, D.~Dhar, A.~Gadde, G.~Mandal, S.~Minwalla, and G.~Sreejith for useful discussions. We gratefully acknowledge generous allocation of computing resources by the Department of Theoretical Physics (DTP) of the Tata Institute of Fundamental Research (TIFR), and related technical assistance from K. Ghadiali and A. Salve. The work of SK was supported at the TIFR by a graduate fellowship from DAE, India. In addition, SK was supported at the ICTS-TIFR by a postdoctoral fellowship from DAE, India in the final stages of this work. KD was supported at the TIFR by DAE, India, and in part by a J.C. Bose Fellowship (JCB/2020/000047) of SERB, DST India, and by
	the Infosys-Chandrasekharan Random Geometry Center
	(TIFR).
	
	\bibliography{references}
	\newpage
		\appendix*
	
	\section{Proof of detailed balance}
	\label{appendix:detail_balance}
	Here, we show that our choices for the initial probability of choosing the initial pivot $\pi_0$ in different cases (as detailed in Sec.~\ref{sec:methodsandobservables3d}), together with local detailed balance conditions imposed at each intermediate step, satisfy the overall detailed balance condition that is a sufficient condition for ensuring that the equilibrium Gibbs distribution is correctly sampled in long-enough simulations. 
The overall detailed balance condition may be stated as follows: If a configuration $C_i$ of weight $W(C_i)$ is updated to another configuration $C_f$ of weight $W(C_f)$ with a probability $P(C_i\to C_f)$ by one step of the worm algorithm, and if the corresponding probability for the backward step starting from configuration $C_f$ is $P(C_f\to C_i)$, then these probabilities must satisfy
	\begin{equation}
		\label{eq:global detailed balance}
		P(C_i\to C_f)W(C_i)=P(C_f\to C_i)W(C_f) \; .
	\end{equation}
Both in the fully-packed case, and in the case with a nonzero fugacity $f_{1/2}$ for half-integer charges, the probability $P(C_i\to C_f)$ of a forward worm construction consisting of $n$ intermediate steps takes the product form
	\begin{align}
		\label{eq:fwdprob}
		&P(C_i\to C_f)=\nonumber \\
		& P_{\pi_0}(C_i)T^{\pi_0}_{e_0 e_1}P_{e_1 \pi_1}(C_i^{(1)})T^{\pi_1}_{e_1 e_2}\cdots P_{e_n \pi_{n}}(C_i^{(n)})T^{\pi_{n}}_{e_{n}e_{n+1}} \; .
	\end{align}
Here, $P_{\pi_0}(C_i)$ is the probability of choosing the first pivot site $\pi_0$ to start a worm construction from configuration $C_i$. Likewise, $P_{e_k \pi_k}(C_i^{(k)})$ denotes the probability of reaching pivot $\pi_k$ via entry  $e_k$ in this forward worm construction, where $C_i^{(k)}$ is the corresponding intermediate configuration with the worm head located at $e_k$. In the above expression, the worm terminates after the $n$th step by exiting pivot $\pi_n$ via exit $e_{n+1}$, and the final configuration $C_f$ can be consistently expressed in terms of the initial configuration as $C_f=C_i^{(n+1)}$. We emphasize that the above decomposition is equally valid both in the fully-packed case and in the case with a nonzero $f_{1/2}$ since the latter case also fits into this description with the understanding that the choices for the entry and exit include the off-lattice entry and exit described in Sec.~\ref{subsec:generalizedMC}.

 We now proceed as follows: Given that a pivot $\pi_k$ is entered from entry $e_k$, the exit $x_k$ through which the worm head leaves the pivot is chosen from the allowed possibilities using probabilities $T^{\pi_k}_{e_k x_k}$ that satisfy the local detailed balance equation (Eq.~\ref{eq:detailed_balance}). As explained in Sec.~\ref{subsec:generalizedMC}, the exit $x_k$ serves as the entry site $e_{k+1}$ of the next pivot at each intermediate step. We indicate this by writing $T^{\pi_k}_{e_k e_{k+1}}$ in place of $T^{\pi_k}_{e_k x_k}$ in Eq.~\ref{eq:fwdprob}.

	\iffalse Here $C_i^{k}$ denotes the intermediate configuration of a forward worm construction, when a worm head moves $k$ steps starting from a configuration $C_i$. The worm ends at the $n+1$'th step leading to the final configuration $C_f\equiv C_i^{n+1}$. The worm head in an intermediate configuration $C_i^{k}$ is located at $e_k$, which serve as the entry site for the $k+1$th step of worm movement. Corresponding pivot site $\pi_k$ is chosen randomly among the sites linked with $e_k$ by a dimer or loop segments except $\pi_{k-1}$. The probability of choosing $\pi_k$ in the intermediate configuration $C_i^{k}$ is denoted as $P_{e_k \pi_k}(C_i^{k})$. \fi
	
	Although the notation employed above was chosen to be natural from the point of view of the forward worm construction, we can nevertheless write the corresponding probability for the backward worm construction using this notation as follows:
	\begin{align}
		&P(C_f\to C_i) = \nonumber \\ &P_{\pi_n}(C_f)T^{\pi_n}_{e_{n+1} e_n}P_{e_n \pi_{n-1}}(C_i^{(n)})T^{\pi_{n-1}}_{e_n e_{n-1}}\cdots P_{e_1 \pi_{0}}(C_i^{(1)})T^{\pi_{0}}_{e_1 e_0}.
	\end{align}
Next, we note that at any intermediate step corresponding to configuration $C_i^{(k)}$, the probability of choosing $\pi_k$ in the forward worm construction ($C_i\to C_f$) equals the probability of choosing $\pi_{k-1}$ in the backward worm construction ($C_f\to C_i$), {\em i.e.},
	\begin{equation}
		\label{eq:choose pivot}
		P_{e_k \pi_k}(C_i^{(k)})|_{C_i\to C_f} = P_{e_k \pi_{k-1}}(C_i^{(k)})|_{C_f\to C_i}
	\end{equation}
	for all $k=1,2\cdots n$. 
	
	We denote the Boltzmann weight associated with the configuration $C_i^{(k)}$ corresponding to the worm head at $e_k$ by $W(C_i^{(k)})$. In a forward worm construction, $e_k$ is reached by exiting the worm head from $\pi_{k-1}$; therefore, in our earlier notation in Sec.~\ref{subsec:generalizedMC}, $W(C_i^{(k)}) = w_{e_k}^{\pi_{k-1}}$. Similarly, in the backward worm construction starting from configuration $C_f\equiv C_i^{(n+1)}$, the configuration $C_i^{(k)}$ appears when the worm head reaches $e_k$ after exiting pivot $\pi_{k+1}$. Hence, in that case, $w^{\pi_{k+1}}_{e_k}\equiv W(C_i^{(k)})$. Using this unified notation, we rewrite the local detailed balance condition at any intermediate step as
	\begin{eqnarray}
		\label{eq:local detailed balance}
		W(C_i^{(k)})T^{\pi_k}_{e_k e_{k+1}} = W(C_i^{(k+1)})T^{\pi_k}_{e_{k+1}e_k}.
	\end{eqnarray}
	Now, using the above local detailed balance relations Eq.~\ref{eq:choose pivot} and Eq.~\ref{eq:local detailed balance}, we can write the left hand side of Eq.~\ref{eq:global detailed balance} as
	\begin{align}
		&W(C_i)P(C_i\to C_f) \nonumber \\
		&=P_{\pi_0}(C_i) \left(\prod_{1}^{n} P_{e_k \pi_k}(C_i^{(k)}) \right) W(C_i)T^{\pi_0}_{e_0 e_1} T^{\pi_1}_{e_1 e_2} \cdots T^{\pi_n}_{e_n e_{n+1}} \nonumber\\
		&= P_{\pi_0}(C_i) \left(\prod_{1}^{n} P_{e_k \pi_{k}}(C_i^{(k)}) \right) T^{\pi_0}_{e_1 e_0} W(C_i^{(1)})  T^{\pi_1}_{e_1 e_2} \cdots T^{\pi_n}_{e_n e_{n+1}}\nonumber\\
		&= P_{\pi_0}(C_i) \left(\prod_{1}^{n} P_{e_k \pi_{k}}(C_i^{(k)}) \right) T^{\pi_0}_{e_1 e_0} T^{\pi_1}_{e_2 e_1} W(C_i^{(2)}) \cdots T^{\pi_n}_{e_n e_{n+1}}\nonumber\\
		& \qquad\qquad \qquad \qquad \qquad \qquad \qquad  \vdots \nonumber \\
		&= P_{\pi_0}(C_i) \left(\prod_{1}^{n} P_{e_k \pi_{k}}(C_i^{(k)}) \right) T^{\pi_0}_{e_1 e_0} T^{\pi_1}_{e_2 e_1} \cdots W(C_i^{(n)})T^{\pi_n}_{e_n e_{n+1}}\nonumber\\
		&= P_{\pi_0}(C_i) \left(\prod_{1}^{n} P_{e_k \pi_{k-1}}(C_i^{(k)}) \right) T^{\pi_0}_{e_1 e_0} T^{\pi_1}_{e_2 e_1} \cdots T^{\pi_n}_{e_{n+1} e_n}W(C_f)\nonumber\\
		&= \frac{P_{\pi_0}(C_i)}{P_{\pi_n}(C_f)} W(C_f)P(C_f\to C_i).
	\end{align}
	Therefore, to satisfy the global detailed balance, we must choose the probability of
	selecting the initial pivot such that
	\begin{equation}
		P_{\pi_0}(C_i)=P_{\pi_n}(C_f)\; ,
	\end{equation}
	{\em i.e.}, the probability of choosing $\pi_0$ for the first pivot of a forward worm construction starting with configuration $C_i$ should be equal to the probability of choosing $\pi_n$ for the first pivot in the corresponding reverse worm construction starting from $C_f$. Our procedure for picking an allowed pivot starting from any configuration implies that these probabilities satisfy $P_{\pi_0}(C_i)= P_{\pi_n}(C_f) = 1/2$ independent of configuration. As a result, the above condition is trivially satisfied and this concludes the proof of detailed balance for both the fully-packed case and for the case with nonzero $f_{1/2}$.
	
\end{document}